\documentclass{aastex}

\usepackage{graphicx}

\usepackage{emulateapj5}
\usepackage{times}

\usepackage{rotating}

\makeatletter

\newenvironment{inlinefigure}{%
\def\@captype{figure}%
\noindent\begin{minipage}{0.999\linewidth}\begin{center}}
{\end{center}\end{minipage}\smallskip}
\makeatother

\usepackage{mathptmx}

\begin{document}

\slugcomment{submitted to {\em The Astrophysical Journal}}

\title{AGN Driven Weather and Multiphase Gas in the Core of the NGC 5044 Galaxy Group}

\author{Laurence P. David${^1}$, Ewan O'Sullivan${^2}$, Christine Jones${^1}$, Simona Giacintucci${^1}$, 
Jan Vrtilek${^1}$, Somak Raychaudhury${^2}$, Paul Nulsen${^1}$, William Forman${^1}$, Ming Sun${^3}$ \& Megan Donahue${^4}$}

\medskip

\affil{${^1}$Harvard-Smithsonian Center for Astrophysics, 60 Garden St.,
Cambridge, MA 02138}
\affil{${^2}$School of Physics and Astronomy, University of Birmingham, Birmingham B15 2TT}
\affil{${^3}$Department of Astronomy, University of Virginia, P.O. Box 400325, Charlottesville, VA 22901}
\affil{${^4}$Department of Physics and Astronomy, Michigan State University, East Lansing, MI 48824}

\shorttitle{\emph NGC 5044}

\begin{abstract}
A deep {\it Chandra} observation of the X-ray bright group, NGC 5044,
shows that the central region of this group has been strongly
perturbed by repeated AGN outbursts.  These recent AGN outbursts
have produced many small X-ray cavities, cool filaments and
cold fronts. We find a correlation between
the coolest X-ray emitting gas and the morphology
of the H$\alpha$ filaments. The H$\alpha$ filaments are oriented
in the direction of the X-ray cavities, suggesting 
that the warm gas responsible for the H$\alpha$ emission originated 
near the center of NGC 5044 and was dredged up behind the buoyant, 
AGN-inflated X-ray cavities.  A detailed spectroscopic analysis shows that 
the central region of NGC 5044 contains spatially varying
amounts of multiphase gas.  The regions with the most inhomogeneous 
gas temperature distribution tend to correlate with the 
extended 235~MHz and 610~MHz radio emission detected by the GMRT.  
This may result from gas entrainment within the radio emitting 
plasma or mixing of different temperature gas in the regions
surrounding the radio emitting plasma by AGN induced turbulence.
Accounting for the effects of multiphase gas, we find that 
the abundance of heavy elements is fairly uniform 
within the central 100~kpc, with abundances of 60-80\% 
solar for all elements except oxygen, which has a 
significantly sub-solar abundance.
In the absence of continued AGN outbursts, the gas in the 
center of NGC 5044 should attain a more homogeneous
distribution of gas temperature through the dissipation
of turbulent kinetic energy and heat conduction in 
approximately $10^8$~yr.  The presence of multiphase gas in
NGC 5044 indicates that the time between recent AGN 
outbursts has been less than $\sim 10^8$~yr.

\end{abstract}


\section{Introduction}

Chandra and XMM-Newton observations have shown that there is a strong
feedback mechanism between the central AGN and the energetics of the
hot gas in groups and clusters of galaxies
(e.g., Peterson \& Fabian 2006 and McNamara \& Nulsen 2007).
AGN outbursts in the central dominant galaxy in groups and
clusters, which are themselves fueled by the accretion of cooling gas,
can produce shocks, buoyant cavities and sound waves, all of which lead
to re-heating of the cooling gas.  This AGN-cooling flow feedback mechanism
is also an important process in galaxy formation and
can explain the observed correlation between bulge mass and central 
black hole mass (e.g., Gebhardt et al. 2000) and the cut-off in the number 
of massive galaxies (Croton et al. 2006).  

AGN outbursts can have a much more significant impact on the 
energetics of the hot gas in groups compared to that in richer
clusters, due to their shallower 
potential wells.  In David et al. (2009), hereafter Paper I, we presented 
the results of an 80~ksec Chandra observation of the X-ray bright group, 
NGC 5044, and showed that the central region 
of this group has been perturbed by several recent AGN outbursts and also by
motion of the central galaxy relative to the surrounding intragroup
gas. GMRT observations at 235~MHz and 610~MHz
show direct evidence for uplifting of low entropy gas from the center
of NGC 5044 behind buoyant, AGN-inflated X-ray cavities. Most of the
X-ray cavities in NGC 5044, however, remain radio quiet in our 
GMRT observations. The roughly isotropic distribution of the small
radio quiet cavities suggests that the group weather has a significant 
impact on the dynamics of these bubbles as they buoyantly rise.
We also showed in Paper I that the total mechanical power of 
the small radio quiet cavities is sufficient to suppress about
one-half of the total radiative cooling within the central 10~kpc.  
In this paper, we discuss how AGN driven turbulence has produced a 
multiphase medium in the center of NGC 5044 and how and this impacts 
estimates of the abundance and distribution of heavy elements.

This paper is organized in the following manner. Details of the ACIS data analysis
are described in $\S 2$. Section 3 presents temperature and abundance maps
for the center of NGC 5044 and $\S 4$ contains a discussion about systematic
effects in the derivation of elemental abundances when fitting emission
from multiphase gas with non-solar abundance ratios.  In $\S 5$
we show how abundance profiles for NGC 5044 depend on the assumed spectral model.
Section 6 presents a comparison between XMM-Newton and Chandra derived abundances.
Correlations between the X-ray and H~$/alpha$ emission are discussed
in $\S 7$.  The implications of our results are discussed in $\S 8$ and 
$\S 9$ contains a brief summary of our main results.

\section{Data Analysis}

NGC 5044 was observed by {\it Chandra} on March 7, 2008 (ObsId 9399) for
a total of 82710~sec. All data analysis was performed with CIAO 4.2 and
CALDB 4.2.1. Since the X-ray emission from NGC 5044 completely fills the S3 chip,
the back-side illuminated chip S1 was used to screen for background flares.  Light 
curves were made in the 2.5-7.0~keV and 9.0-12.0~keV energy bands for the diffuse 
emission on S1.  No background flares exceeding a count rate threshold of 20\% above 
the quiescent background count rate were detected during the observation.
Background images were generated from the standard set of blank
sky images in the Chandra CALDB and the exposure times in the background
images were adjusted to yield the same 9.0-12.0~keV count rates as
in the NGC 5044 data (see Paper I for more details about the data analysis).
The GMRT data for NGC 5044 was obtained as part of a low frequency radio survey
of elliptical-dominated galaxy groups (see Giacintucci et al. 2010;
Gitti el at. 2010).
Throughout this paper we use a luminosity distance
of $D_L=38.8$~Mpc for NGC 5044, which corresponds to $1^{\prime\prime}=185$~pc.

\section{Temperature and Abundance Maps}

A temperature map covering the ACIS-S3 chip (approximately 90~kpc
on a side) is shown in Figure 1.  The temperature map was 
derived using the same method described in O'Sullivan et al. (2005). All 
spectra were fit to an absorbed {\it apec} model with the absorption fixed to the
galactic value ($N_H(gal)=4.94 \times 10^{20}$~cm$^{-2}$)
and the temperature, abundance and normalization treated
as free parameters. Regions were chosen to include at least 1600 net counts.
The middle panel of Figure 1 shows the best-fit temperature and the left and right
hand panels show the 90\% lower and upper limits on the temperature.  
The temperature map shown in Figure 1 is very similar to the temperature 
map presented in Paper I, which was derived from the energy centroid 
of the blended Fe-L lines.  Figure 1 shows that there is a great deal 
of structure in the temperature map and that the coolest gas 
extends toward the SE.  As discussed in Paper I, all of the bright 
X-ray filaments contain cooler gas compared to the surrounding regions.  
A more in depth discussion of the features in the NGC 5044 temperature map
is given in Paper I.

The corresponding abundance map for NGC 5044 is shown in Figure 2.
All 3 panels in Figure 2 show the presence of cloud-like and filament-like 
regions of high abundance gas.  There is also a large extended region of 
very low abundance gas in the SE corner of the S3 chip.  Most of the high 
abundance clouds are 
located at larger radii than the central X-ray cavities (see Figure 3), but 
interior to the SE cold front noted in Paper I (see Figure 4).
The high abundance filaments are primarily located at larger 
radii than the SE cold front.
Except for the spatial coincidence between the high abundance cloud 
toward the NE and an X-ray bright filament, there are no significant
correlations between the high abundance regions, features
in the ACIS image, the unsharp masked image, or the temperature map.

As discussed in Paper I, the 235~MHz radio emission passes through the
southern X-ray cavity, bends to the west just behind the SE cold front 
and then undergoes another sharp bend toward the south, at the location 
of the SE cold front.  Figure  4 shows that the 235MHz radio 
emission passes through regions with the lowest abundances
and threads the region between the high abundance clouds.
The lowest abundance gas is located in the SW corner of S3, at
the same position as the detached 235~MHz radio lobe. 

To determine the accuracy of the method used for deriving the 
abundances shown in Figure 2, we extracted spectra from 7 
regions with super solar abundances and 
5 regions with sub solar abundances (see Figure 5).
These spectra were then fit to a single {\it apec} model in the
0.5-3.0~keV energy band with the absorption fixed at the 
galactic value (see the results in Table 1).  These results are 
in good agreement with the trends seen in the abundance map and 
show that the abundances differ by a factor of approximately 3 
between the high and low abundance regions.  

\section{Systematic Effects in Measuring Abundances of Heavy Elements}

Care must always be taken when estimating abundances for 
gas which is likely multiphase (e.g., Buote et al. 1999; Rasia et al. 2008).  
In general, abundance estimates derived from fitting CCD resolution 
spectra of multiphase gas are highly 
model-dependent.  Abundances are usually underestimated when fitting the
emission from multi-phase gas with temperatures of $\sim$1~keV 
with single temperature models.
The abundance map shown in Figure 2 is based on fitting the projected
spectra to an absorbed {\it apec} model with the absorption fixed at 
the galactic value.  In addition to the presence of multiphase gas, the
derived abundances can also be affected by unresolved LMXBs,
non-solar ratios of heavy elements, excess absorption and projection effects. 

\subsection{Unresolved LMXBs}

In paper I, we found that the power-law emission from unresolved LMXBs can
affect the derived spectral properties in the central region of NGC 5044.
We also derived the ratio between the X-ray luminosity of the LMXBs and 
the K-band luminosity of NGC 5044 in Paper I. Using this ratio, and the 2MASS 
image of NGC 5044, we re-fit the spectra extracted from the high and
low abundance regions shown in Figure 5 with an absorbed {\it apec} plus 
power-law model with the normalization of the power-law component frozen 
at the value derived from the K-band luminosity within each region.  
We found that the improvement in the quality of the fits is statistically insignificant 
(based on a F-test) and that there is no significant effect on the derived abundances.  
This is mainly due to the use of a soft energy band (0.5-3.0~keV) 
in our spectral analysis.  Paper I showed that unresolved emission from LMXBs 
primarily affects the spectroscopic results
when higher energy photons are included in the analysis.

\subsection{Non-Solar Abundance Ratios}

We next fit the set of spectra to an absorbed {\it apec} model with free absorption, 
an absorbed {\it vapec} model with free O, Mg, Ne, Si, S  and Fe abundances (freeing
up one element at a time) and an absorbed two temperature {\it vapec} model.
The single greatest improvement in the quality of the spectral fits
(based on a F-test) was obtained by fitting the spectra with an
absorbed {\it vapec} model with the absorption fixed at the galactic value and 
O and Fe allowed to vary independently.
The results of this spectral analysis are shown in Table 2.
This table shows that for 6 of the 7 high abundance regions, there is a
statistically significant improvement in the quality of the spectral fits.  
For the low abundance regions, the quality of the fits is only improved 
in 1 out of 5 regions.  While the derived Fe abundances in the high abundance regions
are still significantly greater than the Fe abundances in the low abundance
regions (compare Tables 1 and 2), the ratio of the average Fe abundance 
between the high and low abundance regions is significantly reduced 
by allowing O and Fe to vary independently.
Figure 6 shows that the high and low abundance regions also have different O/Fe ratios.
The high abundance regions tend to have low O/Fe ratios, while
the low abundance regions have O/Fe ratios mostly consistent with the
solar ratio (which is why the low abundance regions are well fit 
with an {\it apec} model).

To demonstrate the dependence of the abundance, Z, derived from 
fitting an {\it apec} model to gas with a non-solar O/Fe ratio, we simulated
a set of 1~keV spectra using a {\it vapec} model with a solar abundance of
Fe and a range of O/Fe values.  These spectra were then fit with an {\it apec} model.  
Figure 7 shows that the best-fit abundance derived from an {\it apec} model 
can be significantly overestimated when fitting the emission from gas with an
intrinsically low O/Fe ratio.  By excising the O emission in the spectral analysis,
and only fitting the data in the 0.8-3.0~keV energy band,
we find that the abundance derived from an {\it apec} model is consistent with 
the Fe abundance derived from a {\it vapec} model (see Table 3).
Freeing up additional heavy elements in the {\it vapec} model (i.e., Mg, Ne, Si 
and S) did not produce a statistically significant improvement in the spectral 
analysis or alter the best-fit O and Fe abundances given in Table 2.  

\subsection{Excess Absorption}

It is possible that the low O/Fe ratios derived in the high 
abundance regions are due to excess absorption which
would preferentially absorb the O emission-lines. We therefore generated
a set of simulated {\it vapec} spectra with solar O/Fe ratios and
a range of excess absorption.  These spectra were then fit to 
a {\it vapec} model with the O and Fe abundances treated as free parameters and
the absorption fixed at the galactic value.  The sensitivity of the derived 
O/Fe ratio as a function of excess absorption is shown in Figure 8.
This figure shows that at least $5 \times 10^{20}$~cm$^{-2}$ of excess absorption 
is required to produce the low O/Fe ratios observed in the high 
abundance regions.  Table 4 shows the results of fitting the sample spectra
to a {\it vapec} model with $N_H$ and the O and Fe abundances treated
as free parameters.  Based on a F-test, there is
very little statistical improvement in the quality of the fits 
compared to that obtained with $N_H$ fixed at the galactic value.
In addition, only 2 of the 12 regions have a best-fit excess $N_H$ 
of more than $1.0 \times 10^{20}$~cm$^{-2}$ at the 90\% confidence limit.
There is also no correlation between the O/Fe ratio and the 
best-fit excess $N_H$ among these regions.

The low energy QE of the ACIS detectors has continued
to decrease during the course of the {\it Chandra} mission due to 
out gassed material condensing 
on the cold ACIS filters. Our analysis uses the latest update
to the ACIS contamination model which was released in CALDB 4.2 on 
December 15, 2009, however, systematic uncertainties in the ACIS contamination
model could affect our derived O abundances. In addition to the
energy dependence of the absorbing material, the ACIS-S contamination 
model also includes a time-dependence and a dependence on chipy
(parallel to the read-out direction). To estimate the 
systematic uncertainty in the CALDB 4.2 version of the
ACIS contamination model, we extracted a spectrum from the 2009 observation
of the Coma cluster (which has a very low galactic column density 
of $N_H=8.5 \times 10^{19}$) spanning the same range in chipy as that
covered by the high abundance clouds in the NGC 5044 observation.
The spectrum was then fit to an absorbed {\it apec} model
with the absorption treated as a free parameter.  We obtained
a best-fit $N_H$ of $5.5 \times 10^{19}$ with a 90\% upper
limit of $N_H=1.3 \times 10^{20}$.  Thus, excess absorption due
to systematic uncertainties in the ACIS contamination model 
should be less than $4.5 \times 10^{19}$ around the time of the
NGC 5044 observation. Such a small systematic uncertainty cannot
account for the low observed values of the O/Fe ratio (see Figure 8).

\subsection{Multiphase Gas}

Table 5 shows the results of fitting the spectra to a 
two-temperature {\it vapec} model with the O and Fe abundances
linked between the two temperature components and the absorption 
fixed at the galactic value.  For regions H1, H7 and L5, the two 
temperature fits were essentially degenerate with respect to the 
single temperature fits (i.e., the values for the two temperatures 
overlap within their uncertainties). 
For these regions, we replicate the results from the single
temperature analysis in Table 5.  Table 5 shows that the improvement in 
the quality of the spectral fits
for 7 of the 9 remaining spectra is highly significant with the
addition of a second temperature component. One obvious difference
between the low and high abundance regions is the 
fraction of the
total flux from the cooler component.  In the high abundance 
regions, the cooler component accounts for approximately 80\% of the total flux, 
while in the low abundance regions, the flux from the cool and hot 
components is more evenly distributed.

Due to the complexity in the abundance map, we cannot perform a simple 
deprojection analysis.  Since we are fitting projected spectra, some of the 
improvement obtained by fitting a two-temperature model to the spectra 
probably arises from the range in gas temperatures along the line-of-sight.  
Using the deprojected temperature and density profiles for NGC 5044,
we calculated the flux arising from gas at different temperatures.
Figure 9 shows the cumulative fraction of the 0.5-3.0~keV flux,
relative to the total flux, arising from gas cooler than a given temperature along
three different projected distances (b=10, 20 and 30~kpc)
from the center of NGC 5044.
Most of the spectral extraction regions shown in Figure 5 are located approximately
20~kpc from the center of NGC 5044. Figure 9 shows that for a projected
distance of b=20~kpc, approximately 20\% of the emission should arise from gas hotter 
than 1~keV due to projection effects.  This is in good agreement
with the $F_c/F_{tot}$ ratios for the high abundance regions (see Table 5), 
indicating that the emission from within the high abundance regions themselves
is nearly single-phased.  The greater fraction of emission from hotter
gas in the low abundance regions indicates that the gas within these regions 
has a greater predominance of multiphase gas. 

The resulting O/Fe ratios and Fe abundances from a two-temperature
spectral analysis are shown in Fig. 10. Fitting the spectra to two-temperature
models lessens the Fe abundance contrast between the regions initially
identified as having high and low abundances based on the spectral analysis
with an {\it apec} model.  From Table 1, the ratio of the average abundance
in the high abundance regions to the average abundance in the low abundance
regions is 2.5. Fitting the spectra with a {\it vapec} model with Fe 
and O treated as free parameters gives an average Fe abundance ratio 
between the high and low abundance regions of 1.6 (see Table 2). The average
Fe abundance contrast between the high and low abundance regions 
is further reduced to 1.3, when the spectra are analyzed with
a two-temperature model (see Table 5). Tables 2 and 5 show
that the primary reason for the reduction in the Fe abundance contrast is
a 52\% increase in the Fe abundance in the low abundance regions
compared to only a 26\% increase in the Fe abundance in the high 
abundance regions, when a second temperature component is added.

Even though the Fe abundance contrast between the low and high
abundance regions is significantly reduced by fitting the spectra
with a two-temperature model, there is still a clear separation between
the high and low abundance regions in the O/Fe and Fe plane
(see Figure 10). On average, the low abundance regions have higher O/Fe ratios than 
those derived for the high abundance regions.  To investigate
the dependence of the O/Fe ratio on the presence of multiphase
gas, we simulated a set of two-temperature spectra and fit the spectra
to a single temperature model. All of the simulated two-temperature
models have a solar O/Fe ratio, a lower temperature of 0.7~keV, an
upper temperature of 1.3~keV and a range in flux ratios
between the two temperature components.  The spectra were then fit to a
single {\it vapec} model with O and Fe allowed to vary independently.
The results are shown in Fig. 11.  This figure shows that the
O/Fe ratio can be overestimated by 30-50\%, due to the presence
of multi-phase gas which can explain the segregation seen 
in the O/Fe and Fe plane shown in Figure 10, if the gas in
the low abundance region has a greater predominance of multiphase gas
and the actual O/Fe ratios are similar between the different 
regions.  These results imply that the apparently low abundance regions
in the abundance map have a more inhomogeneous gas temperature 
distribution compared to the apparently high abundance regions, but that 
the actual Fe abundances and O/Fe ratios are fairly similar.

To further support the claim that the low abundance
regions reside in highly multiphase gas, we fit 
the spectra to a three temperature model with the 
temperatures fixed at 0.7, 1.0 and 1.3~keV and the 
O and Fe abundances linked between the different 
temperature components.  Figure 12 shows the relative 
distribution of the emission measures for the three 
temperature components.  This figure 
clearly shows that the low abundance regions have
a broader temperature distribution of emission measures compared
to that among the high abundance regions.

In summary, there are four reasons to suspect that the 
{\it apec} derived abundance map shown in Figure 4 is more a map
of temperature inhomogeneities rather than a true abundance map.
These are: 1) the higher values of $F_c/F_{tot}$ in the high 
abundance regions compared to the low abundance regions, 
2) the narrower temperature distribution in the emission measures 
within the high abundance regions compared to that in the low abundance 
regions, 3) the greater increase in the derived Fe abundance in the low
abundance regions compared to the high abundance regions 
when a second temperature component is added and 4) the 
smaller values of the O/Fe ratio in the high abundance regions
compared to that in the low abundance regions derived
from a single temperature analysis.

\section{Abundance Profiles}

The model-dependent nature of derived abundances for emission
from multiphase gas with non-solar abundance ratios makes it 
difficult to determine azimuthally 
averaged abundance profiles.  Such information is essential 
for determining the masses and mass ratios of different elements 
and the relative enrichment from Type Ia and Type II supernovae.
To study the sensitivity of derived abundance profiles on the 
assumed spectral model, we produced 3 sets of spectra 
(with 10000, 20000 and 40000 net counts in each spectrum) extracted 
from concentric annuli centered on NGC 5044.
Figure 13 shows the projected Fe abundance profile 
derived by fitting an absorbed {\it vapec} model with all the 
elemental abundances linked to Fe (this is equivalent to fitting 
the spectra to an {\it apec} model).
Also shown in Figure 13 is the projected Fe abundance profile if the 
O and Fe abundances are allowed to vary independently.
The presence of non-solar abundance ratios produces a sharp peak in the Fe 
abundance if all of the elements are linked together.
Allowing the O abundance to vary produces a fairly flat Fe abundance profile
within the central 100~kpc.

To investigate the distribution of other elements,
we also fit the sets of spectra with 20000 and 40000 net counts
to an absorbed {\it vapec} model with O, Mg, Ne, Si, S and Fe
treated as free parameters.  Figure 14 shows the projected Fe, Si
and O abundance profiles and Figure 15 shows the projected Mg and S
abundance profiles.  We do not show the Ne abundances due to 
their large uncertainties. Figures 14 and 15 show, that except for O, 
the other elements have essentially solar ratios.  This explains
why the spectral analysis of the low and high abundance regions
analyzed in $\S 6$ did not improve significantly when elements 
other than O were allowed to vary independently of Fe.

The deprojected Fe, O and Si abundance profiles derived from fitting
single and two-temperature models are shown in Figures 16 and 17. The spectra 
were deprojected using the XSPEC task {\it projct}.  These figures show
that there is little difference between the projected and deprojected Fe and 
Si abundance profiles, since they are both essentially flat within the
central 100~kpc. The O/Fe ratio is also significantly less than the 
solar ratio for both the single and two-temperature deprojections.

\section{Comparison with XMM-Newton EPIC and RGS Results}

Buote et al. (2003) derived azimuthally averaged abundance profiles
for NGC 5044 from a joint analysis of an earlier 20~ksec {\it Chandra} 
observation and an {\it XMM-Newton} observation.  They found significant
improvements in the spectral analysis when the data were fit with
two-temperature or differential emission measure models compared to 
single temperature models and discussed the sensitivity of the derived
abundances on the presence of multiphase gas.  Most of their abundance 
profiles have a peak abundance between 10-30~kpc with lower values both interior
and exterior to this region.  The peak abundances in their profiles
correspond to the ring of high abundance clouds seen in the
abundance map (see Figure 2).  The peak Fe abundances in Buote et al.
are approximately 0.8 and 1.2 solar in the projected and deprojected analysis,
which are in good agreement with the results presented here.

Tamura et al. (2003) presented an analysis of the {\it XMM-Newton} Reflection
Grating Spectrometer (RGS) observation of NGC 5044 and found that the
emission within the central $2^{\prime}$ (22~kpc) was well fit 
by a two-temperature model with temperatures of 0.7 and 1.1~keV
(consistent with the range in temperatures along the line-of-sight),
an Fe abundance of 0.55 and an oxygen abundance of 0.25 (relative to the
abundance table in Anders \& Grevesse 1989). Converting these values
using the abundance table in Grevesse \& Sauval (1998) gives Fe=0.81 and O/Fe=0.39.
These values for the Fe abundance and O/Fe ratio fall between the values
for the high and low abundance regions in Figure 10.

\section{H$\alpha$ emission}

A narrow-band H$\alpha$ image of NGC 5044 was obtained 
on April 10, 2010 with the 4.1~m Southern Observatory for Astrophysical 
Research (SOAR) telescope using the SOAR 
Optical Imager (SOI). Two CTIO narrow-band filters were used:
6649/76 for the H$\alpha$+[NII] lines and 6520/76 
for the continuum.  Each image was reduced using the standard 
procedures in the IRAF MSCRED package and the spectroscopic 
standard is LTT~3864. More details on the SOI data reduction can be 
found in Sun et al. (2007).

Figure 18 shows the H$\alpha$ emission along with gas temperature 
contours (red) and the locations of the southern, northwestern
and southwestern caviities (blue).
The total H$\alpha$ flux and luminosity is 
$\sim 4.4 \times 10^{-13}$~erg~cm$^{-2}$~s$^{-1}$
and $\rm{L_{H\alpha}} \sim 7.9 \times 10^{40}$~erg~s$^{-1}$. 
The H$\alpha$ emission is primarily extended in a north-south
direction, as previously reported by Macchetto et al. (1996).
Figure 18 shows that there are many emission-line filaments extending 
at least 10~kpc from the nucleus.  Most of the filaments extend toward the
NW cavity and along the SW filament, which is located
between the SW and S cavities.  There is a clump of H$\alpha$ emission 
within the NW cavity and some H$\alpha$ emission in the
SW cavity, but very little emission within the southern cavity.  
Both the NW and SW cavities are radio quiet based on our
recent GMRT observations. The brightest and longest of the southern 
H$\alpha$ filaments borders the western edge of the southern cavity.
The temperature contours in Figure 18 show that there is a strong 
correlation between the location of the coolest X-ray emitting gas 
and the H$\alpha$ emission.  Temi et al. (2007) reported
evidence for distributed PAH emission with the same morphology as 
the southern extension of the H${\alpha}$ emission based on a {\it Spitzer} 
observation of NGC 5044.  

As discussed in Paper I, the kinematics of the H${\alpha}$ emitting
gas are inconsistent with an external origin for the gas (Caon et al. 2000)
and are similar to the kinematics of the H${\alpha}$ filaments in 
NGC 1275 (Hatch et al. 2006). We also showed in paper
I that the most likely origin for the H${\alpha}$ emitting gas
was due cooling of the hot X-ray emitting gas since the 
stellar mass loss rate in NGC 5044 is more than an order 
magnitude less than the mass cooling rate.
All of these results are consistent with a scenario in which
the gas responsible for the H${\alpha}$ originates from cooling 
of the hot X-ray emitting gas within the central few kpcs of NGC 5044, 
followed by dredge up behind AGN-inflated, buoyantly rising bubbles. 
The SW H$\alpha$ filament is apparently being dredged up along the outer edge 
of the 235~MHz radio emitting plasma which threads the 
southern cavity (see Paper I).

\section{Discussion}

The multiphase gas in the center of NGC 5044 is likely produced
by AGN induced stirring or turbulence.  If all AGN activity were to cease, 
then the subsequent cascading of the turbulence to smaller scales
and the eventual dissipation of the turbulent kinetic energy into heat, along
with thermal conduction, will eventually produce a more 
homogeneous distribution of gas temperature.
Turbulent kinetic energy is dissipated on approximately the
eddy turnover time of the largest eddy which is given by $t_{eddy}=l/u$, where 
$l$ is the length scale of the largest eddy and $u$ is corresponding 
turbulent velocity.  The dissipation rate of turbulent kinetic 
energy per unit volume is given by (e.g., David \& Nulsen 2008):

\begin{equation}
\Gamma_{diss} = c_2 \rho_g u^2/t_{eddy} = c_2 \rho_g l^2/t_{eddy}^3 
\end{equation}

\noindent
where $c_2$ is a constant of order unity and $\rho_g$ is the gas
density.  We can estimate $t_{eddy}$ by assuming that 
the dissipation rate of turbulent kinetic energy locally balances radiative losses,
i.e., $\Gamma_{diss}=\Delta L_{bol}$, where $\Delta L_{bol}$ is
the bolometric luminosity of the gas per unit volume.
Assuming the turbulence can be characterized by a local mixing length prescription, 
then the size of the largest eddies can be written as $l=\alpha r$, 
where $\alpha$ is a parameter between 0 and 1 and $r$ is the radial 
distance from the cluster center.  Substituting these expressions 
into eq. (1) and solving for $t_{eddy}$ gives:

\begin{equation}
t_{eddy} = \left( { {c_2 \alpha^2 r^2 \rho_g} \over {\Delta L_{bol}} } \right)^{1/3}
\end{equation}

\noindent
Adopting $c_2=0.42$ (based on the discussion in Dennis \& Chandran 2005),
we computed $t_{eddy}$ as a function of radius for NGC 5044 
for $\alpha=0.2$ and $0.5$ (see Figure 19).  This figure shows that without 
further AGN outbursts, the kinetic energy associated with the present 
level of turbulence will be thermalized within approximately $10^8$~yr within the
central 10~kpc.  The required turbulent velocity
to locally balance radiative losses within the central 10~kpc
is $u \sim 20-40$~km~s$^{-1}$ for $\alpha = 0.2-0.5$, which gives a 
turbulent gas pressure less than 1\% of the thermal gas pressure within this region.

A cool cloud embedded in hotter gas will be heated by thermal conduction
at a rate (see the discussion in Voit et al. 2008):

\begin{equation}
H = 4 \pi r_c^2 f_s \kappa (T) \Delta T/r_c
\end{equation}

\noindent
where $r_c$ is the radius of the cool cloud, $\Delta T$ is the temperature 
difference between the cool cloud and the ambient medium, $f_s$ is a 
reduction factor relative to Spitzer conduction and 
$\kappa (T) = 6 \times 10^{-7} T^{5/2}$~erg~s$^{-1}$~K$^{-1}$~cm$^{-1}$.
The time scale for conduction to heat the cool cloud to
the ambient gas temperature at constant pressure is:

\begin{equation}
t_{cond} = { {5 k \Delta T M_c} \over {2 \mu m_p H} }
\end{equation}

\noindent
where $M_c$ is the mass of the cool cloud.  This equation can 
be written as:

\begin{equation}
t_{cond} = { {5 k r_c^2 \rho_c} \over { 6 \mu m_p f_s \kappa(T)} }
\end{equation}

\noindent
where $\rho_c$ is the gas density within the cool cloud.  
The conduction time scale as a function of radius is shown in Fig. 19 assuming a 
20\% difference between the gas temperature in the cool cloud and ambient medium,
or equivalently, a 20\% contrast in density, assuming the clouds are in pressure 
equilibrium, $r_c=1$~kpc and $f_s = 0.1$ or $0.3$. This figure shows that 
$t_{cond}$ is comparable to $t_{eddy}$ within the 
central 10~kpc.  The conduction time scale declines beyond 
the central 10~kpc due to the increasing gas temperature (see the 
temperature profile presented in Paper I)
and the $T^{5/2}$ factor in the heat conduction rate.   

Figure 19 shows that both turbulence and heat conduction will
produce a more homogeneous distribution of gas temperature
within the central 10~kpc in NGC 5044
within approximately $10^8$~yr in the absence of continued AGN 
driven turbulence.  The fact that there is multiphase gas within 
this region implies that the delay between recent AGN outbursts has 
been less than $10^8$~yr.  

While conduction may be efficient at re-heating small cool
clouds beyond the central 10~kpc in NGC 5044 (see Figure 19), conduction
has a negligible affect on preventing the bulk of the central
gas in NGC 5044 from cooling.  Figure 20 shows the value
of $f_s$ required for heat conduction to locally balance radiative losses.
Within the central 20~kpc, where the gas is fairly isothermal,
heat conduction rates in excess of the Spitzer value are required
to suppress cooling.  Beyond 50~kpc, near the 
peak in the temperature profile, heat conduction actually helps to
cool the gas (negative values of $f_s$), since the divergence of the heat
flux is negative in this region.  Thus, there is only
a small range in radii, between 25-50 kpc, where
heat conduction could be important in preventing the gas from
cooling.

\section{Summary}

In Paper I, we showed that the central region of
NGC 5044 has been strongly perturbed by at least three recent AGN outbursts.
By combining GMRT and Chandra data, we found direct evidence for the uplifting 
of low entropy gas from the center of NGC 5044 by AGN-inflated, 
radio emitting X-ray cavities.  The nearly continuous production 
of AGN-inflated bubbles over the past few $10^8$ years 
has probably produced some gas turbulence within the central regions 
of NGC 5044.  This AGN driven turbulence can have a significant 
impact on the distribution of heavy elements and the production
and longevity of multiphase gas in the center of NGC 5044.

In this paper, we have shown that fitting a single temperature thermal
plasma model with the ratios of the elemental abundances fixed
at the solar values (e.g., {\it apec}) can produce significant 
artifacts in abundance maps.  To investigate the origins
of the artifacts in the {\it apec} derived abundance map
for NGC 5044, we extracted 
spectra from several regions with apparently high and low abundances and 
fit these spectra to a variety of spectral models with different combinations
of free parameters.  The single greatest improvement in the spectral analysis was
obtained by fitting the spectra with a {\it vapec} 
model with the O and Fe abundances allowed to vary independently.  
For the apparently high abundance regions, the best-fit 
Fe abundances derived with the {\it vapec} model are significantly 
less than the abundances obtained with the {\it apec} model. 
These high abundance regions also have best-fit values of the O/Fe  
ratio significantly less that the solar value.
For the apparently low abundance regions, the best-fit Fe abundances
derived with the {\it vapec} model are essentially consistent with the abundances
derived with the {\it apec} model.  The low abundance regions have near solar
O/Fe ratios.  Freeing up the abundance of other heavy elements does not 
further improve the spectral fits in any of the extracted spectra, indicating 
that the abundance ratios of the other elements are essentially
consistent with the solar ratios.  This analysis showed that a substantial
portion of the artifacts in the {\it apec} derived
abundance map are due to the enforcement of a solar O/Fe ratio in 
the spectral analysis.  
This result was confirmed by generating a set of simulated {\it vapec} spectra  
with a range of O/Fe values and fitting the data to an {\it apec} model.
We found that the best-fit abundance obtained with an {\it apec} model increases with 
a decreasing O/Fe ratio.

After allowing for non-solar O/Fe values in the spectral analysis,
the next greatest single improvement in the quality of the spectral
fits was obtained by including a second temperature component. 
The addition of a second temperature component produces a slight
increase in the Fe abundance in the apparently high abundance regions
and a much greater increase in the Fe abundance in the apparently low abundance
regions. The O/Fe ratios are consistent between the single and two 
temperature spectral analysis, given the uncertainties. The flux from
the second temperature component in the apparently high abundance regions 
is consistent with that expected from projection affects given the deprojected
temperature profile of NGC 5044.  For the spectra extracted
from the regions with apparently low abundances, the emission 
from the two different temperature components is more evenly distributed.
In addition to fitting the spectra to a two temperature model, 
we also fit the spectra to a three temperature model with the
temperatures fixed at 0.7, 1.0 and 1.3~keV.  This analysis
showed that the emission from the apparently low abundance regions
is more evenly distributed among the different temperature components
compared to the emission from the apparently high abundance regions, 
confirming that inhomogeneities in the temperature distribution 
are partly responsible for producing the artifacts in the {\it apec} 
derived abundance map.  We also found that excess absorption 
and emission from unresolved LMXBs cannot produce the artifacts 
observed in the {\it apec} derived abundance map.

We have shown that deprojected abundance profiles are 
very sensitive to the assumed spectral model.  
A single temperature {\it apec} model produces a deprojected abundance profile 
for NGC 5044 that increases by a factor of approximately
2 from 100 to 10~kpc, and then decreases inward. 
By fitting the same set of spectra with a {\it vapec} model with the O and Fe
abundances allowed to vary independently, we obtain deprojected
O and Fe abundance profiles that are fairly uniform within the central 100~kpc. 
Central dips in abundance profiles are a fairly common occurrence
in cooling flow clusters (e.g., Sanders et al. 2004, Churazov et al. 2004).
Our results show that such dips in the central abundance profile
could be due to non-solar ratios of the heavy elements or
the presence of multiphase gas.
Excluding oxygen, all of the other elements that produce strong
emission lines in 1~keV gas (i.e., Mg, Si and S) have essentially 
flat abundance profiles with near solar ratios within the central 100~kpc.

The O/Fe ratio is a useful diagnostic for determining the 
relative enrichment from Type Ia and Type II
supernovae.  However, estimates of the O/Fe ratio are very sensitive to
the presence of multiphase gas.  By generating a set of 
simulated spectra, we have shown that the O/Fe 
ratio can be overestimated by factors of 2 to 3 when the emission from 
multiphase gas is fit with a single temperature model.  While the uncertainties
on the O/Fe ratios obtained from our spectral analysis 
are fairly large when more than one temperature component
is included in the spectral analysis, on average, we 
find significantly sub-solar values of O/Fe within the 
central  100~kpc of NGC 5044.  This implies a greater relative 
enrichment from Type Ia enrichment compared to solar abundance 
gas which is consistent with previous studies of the central 
regions in groups and clusters (e.g., Finoguenov et al. 2000;2001).

The multiphase gas in NGC 5044 is probably produced by AGN 
induced turbulence.  We have shown that without further AGN outbursts, 
the gas in the center of NGC 5044 should acquire a more
homogeneous distribution of temperature within $10^8$~yr due 
to the dissipation of turbulent kinetic energy
and the effects of thermal conduction.  The fact that there is a 
significant component of multiphase gas in the center of NGC 5044 shows that the
delay between successive AGN outbursts must be less than $10^8$~yr.
We have recently obtained deeper GMRT data for NGC 5044
to better probe the past AGN history in this system and we will 
present the results in a subsequent paper.

\bigskip

We would like to acknowledge Seth Bruch for obtaining the narrow band 
H$\alpha$+[NII] images at the SOAR.  This work was supported by 
grants GO8-9122X and GO0-11003X. EOS acknowledges the support of the 
European Community under the Marie Curie Research Traning Network.

\newpage

\begin{inlinefigure}
\center{\includegraphics*[width=2.00\linewidth,bb=35 284 576 509,clip]{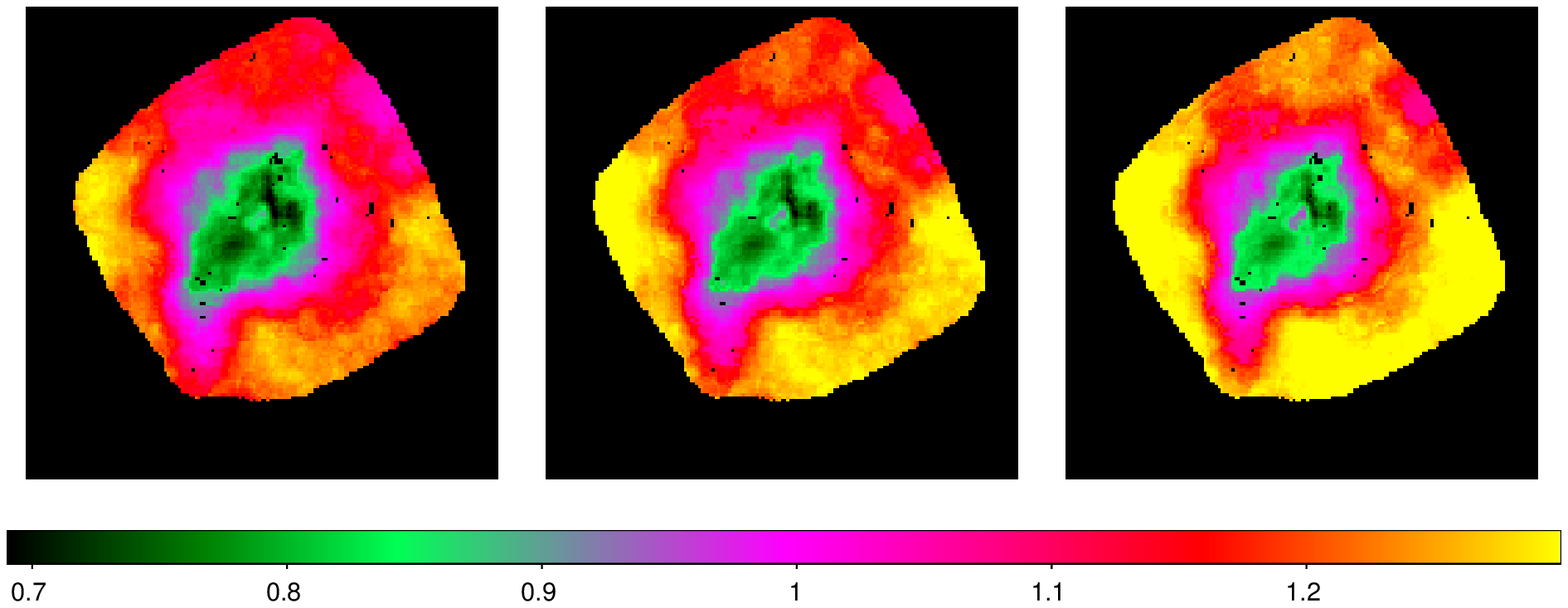}}
\caption{Temperature map covering the ACIS-S3 chip.  The central panel shows
the best-fit temperature and the left and right panels show the 90\% lower and
upper limits on the temperature. The spectra were fit with an absorbed
{\it apec} model with the temperature, abundance and normalization treated
as free parameters and the absorption fixed at the galactic value.}
\end{inlinefigure}

\begin{inlinefigure}
\center{\includegraphics*[width=2.00\linewidth,bb=35 284 576 509,clip]{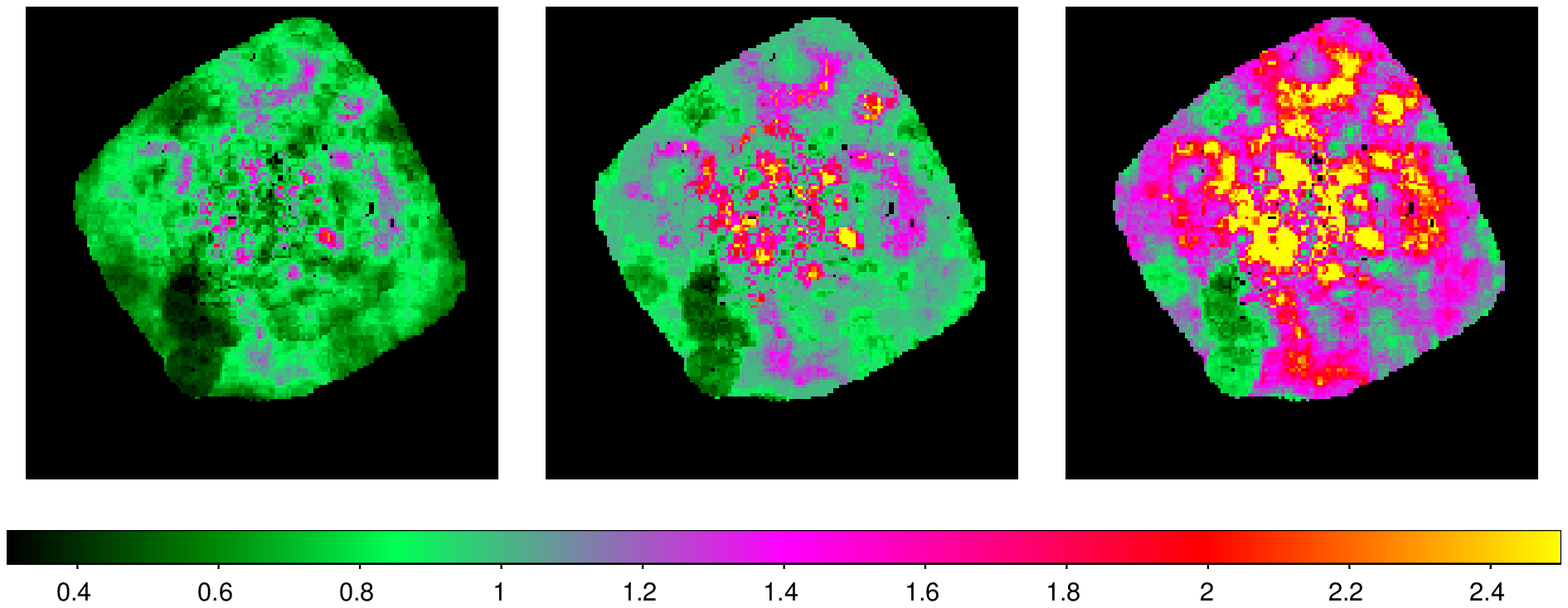}}
\caption{Abundance map covering for the ACIS-S3 chip.  The central panel shows
the best-fit abundance and the left and right panels show the 90\% lower and
upper limits on the abundance. The spectra were fit with an absorbed
{\it apec} model with the temperature, abundance and normalization treated
as free parameters and the absorption fixed at the galactic value.}
\end{inlinefigure}

\newpage

\begin{inlinefigure}
\center{\includegraphics*[width=1.00\linewidth,bb=130 218 484 576,clip]{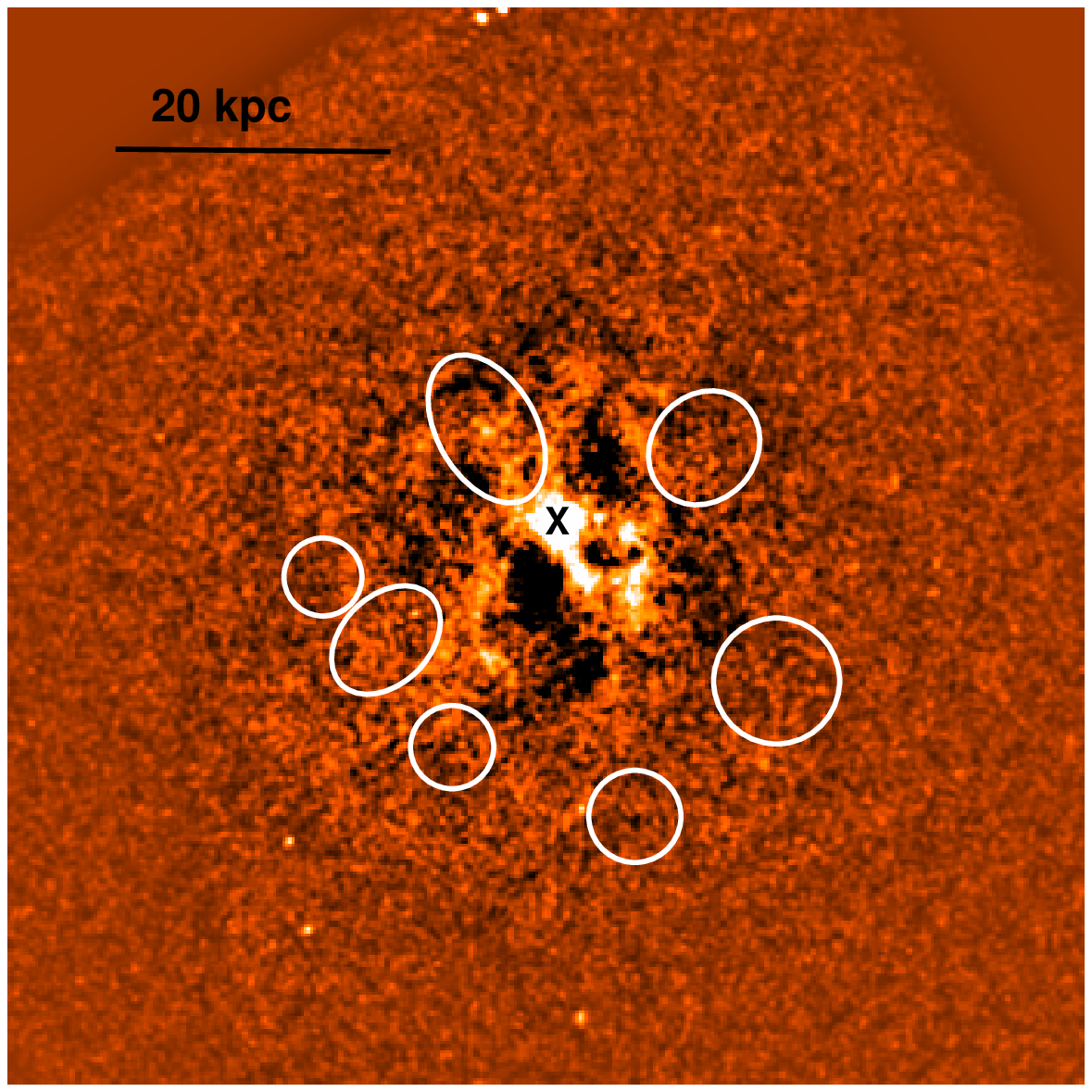}}
\caption{Location of the high abundance clouds detected in the abundance map
highlighted on the unsharp masked image of NGC 5044. The "X" marks the location
of the AGN at the center of NGC 5044.} 
\end{inlinefigure}

\begin{inlinefigure}
\center{\includegraphics*[width=1.00\linewidth,bb=81 163 526 627,clip]{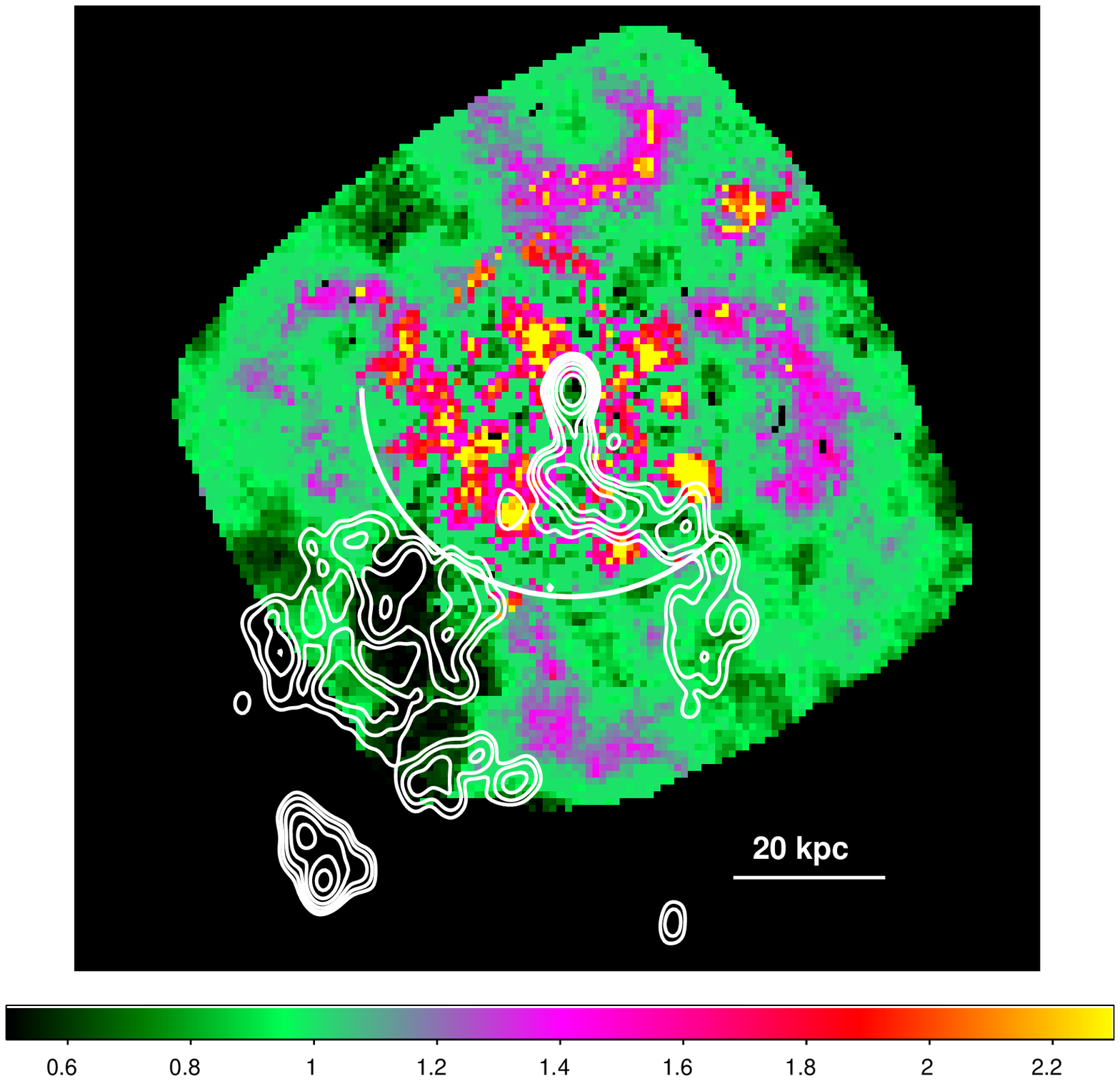}}
\caption{GMRT 235~MHz contours overlayed on the abundance map (central panel of Fig. 4). 
The beam size is 22$^{\prime\prime}$ by 16$^{\prime\prime}$ and the lowest contour is shown at 
$3\sigma = 0.75$~mJy~b$^{-1}$ = 0.2mJy/b.  The white arc shows the location of the SE
cold front (see Paper I).}
\end{inlinefigure}

\begin{inlinefigure}
\center{\includegraphics*[width=1.00\linewidth,bb=90 198 507 580,clip]{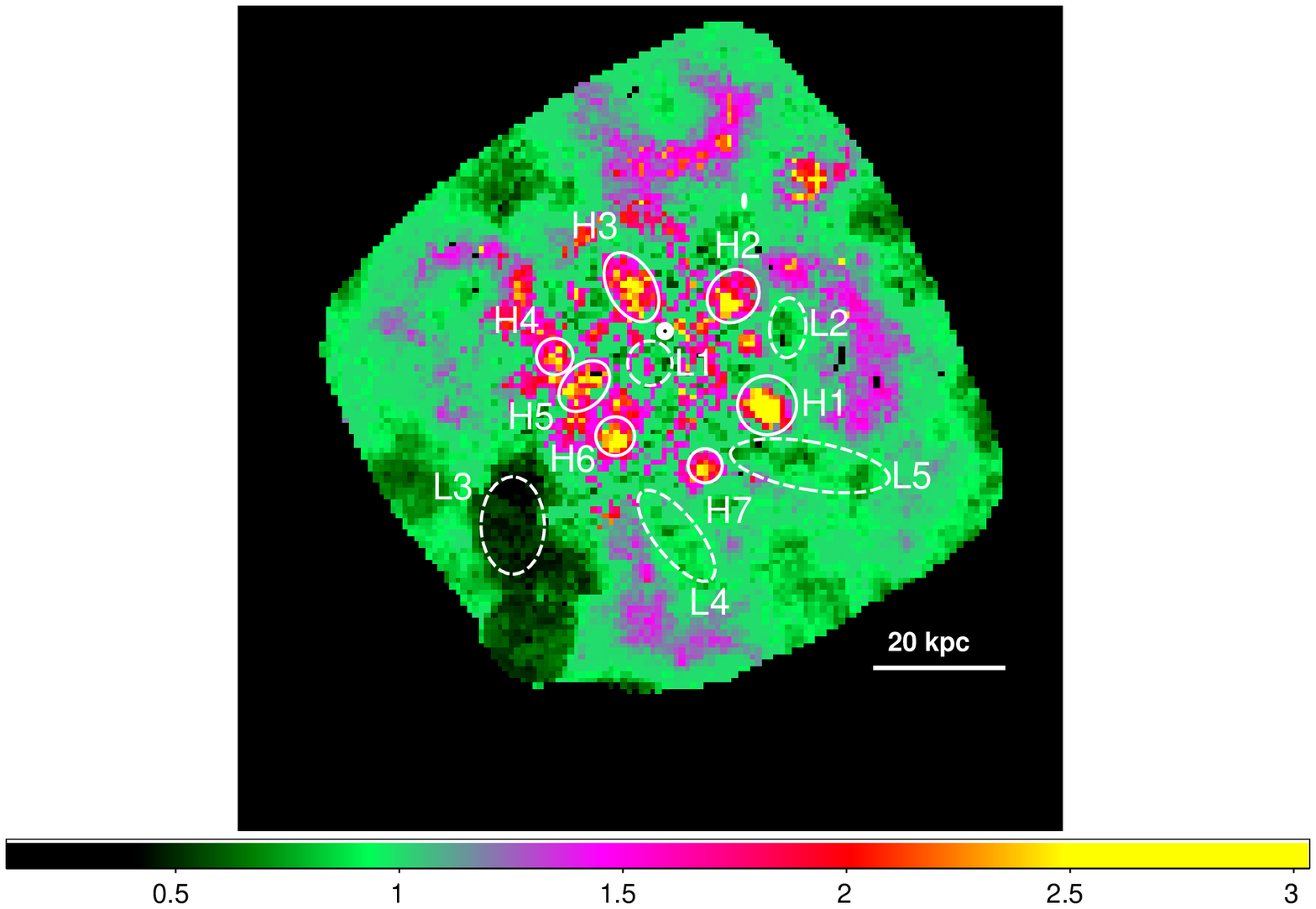}}
\caption{High and low abundance regions used for more in-depth spectral analysis.
The small heavy white circle indicates the location of the AGN at the center of NGC 5044.}
\end{inlinefigure}

\begin{inlinefigure}
\center{\includegraphics*[width=1.00\linewidth,bb=19 145 583 706,clip]{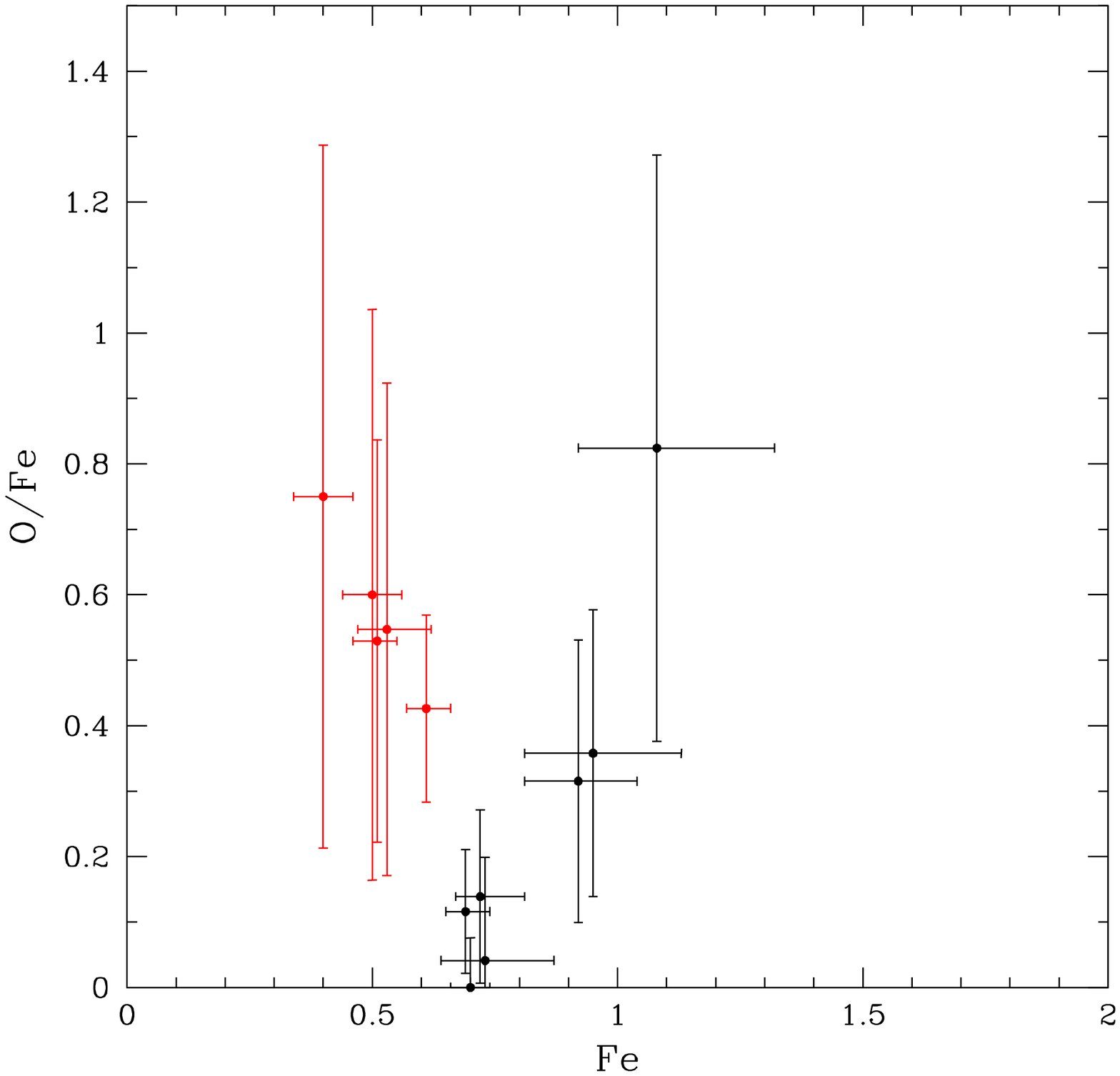}}
\caption{Scatter plot of the O/Fe ratios and Fe abundances for the 
high abundance clouds (black) and low abundance regions (red)
derived from fitting the spectra to a single-temperature {\it vapec} model.  All error
bars are shown at $1 \sigma$.}
\end{inlinefigure}

\begin{inlinefigure}
\center{\includegraphics*[width=1.00\linewidth,bb=25 145 583 706,clip]{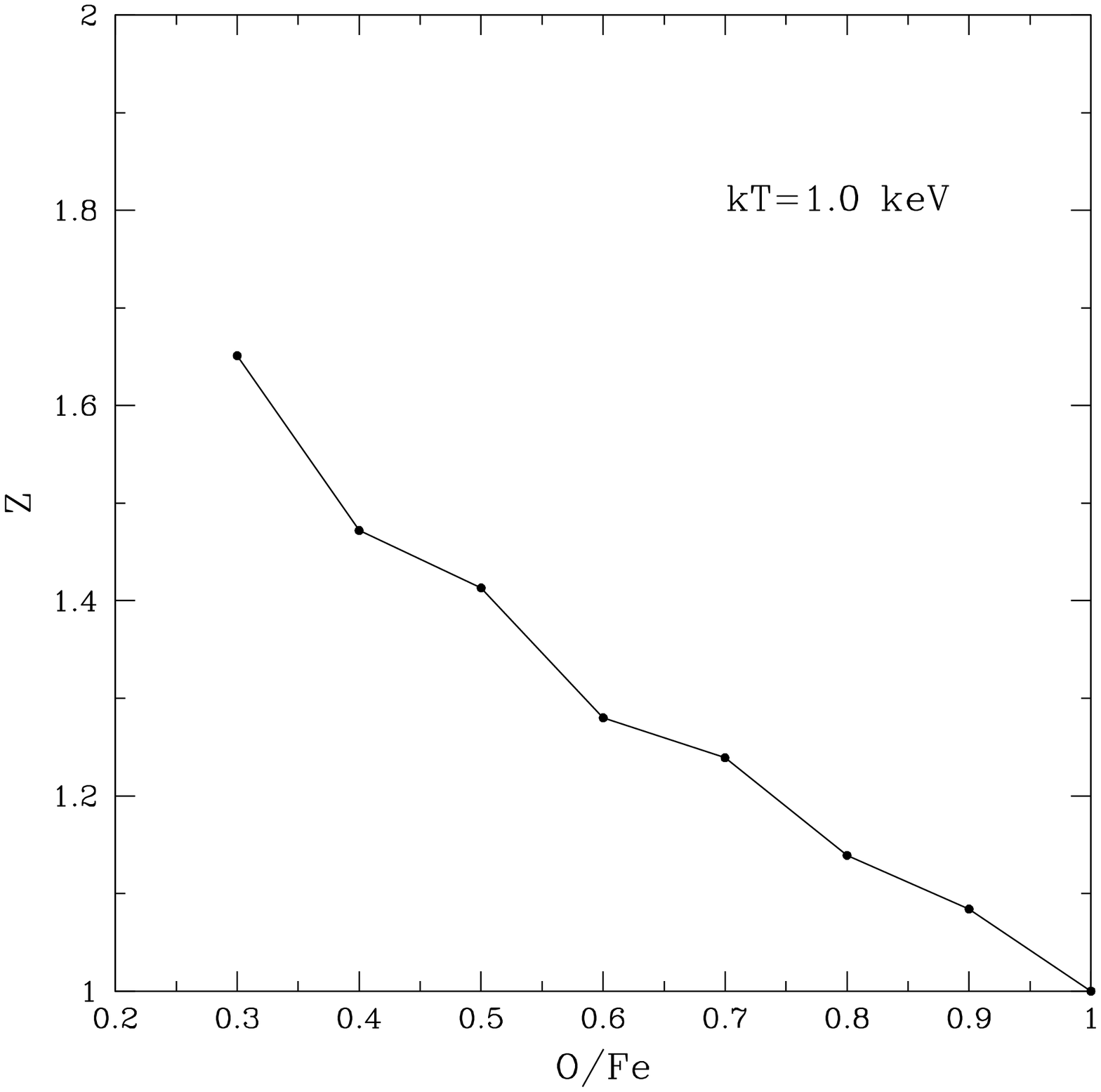}}
\caption{Best-fit abundance (Z) derived from fitting an {\it apec} model to simulated
spectra generated from a {\it vapec} model with a solar Fe abundance and a range of O/Fe values.}
\end{inlinefigure}

\begin{inlinefigure}
\center{\includegraphics*[width=1.00\linewidth,bb=19 145 583 706,clip]{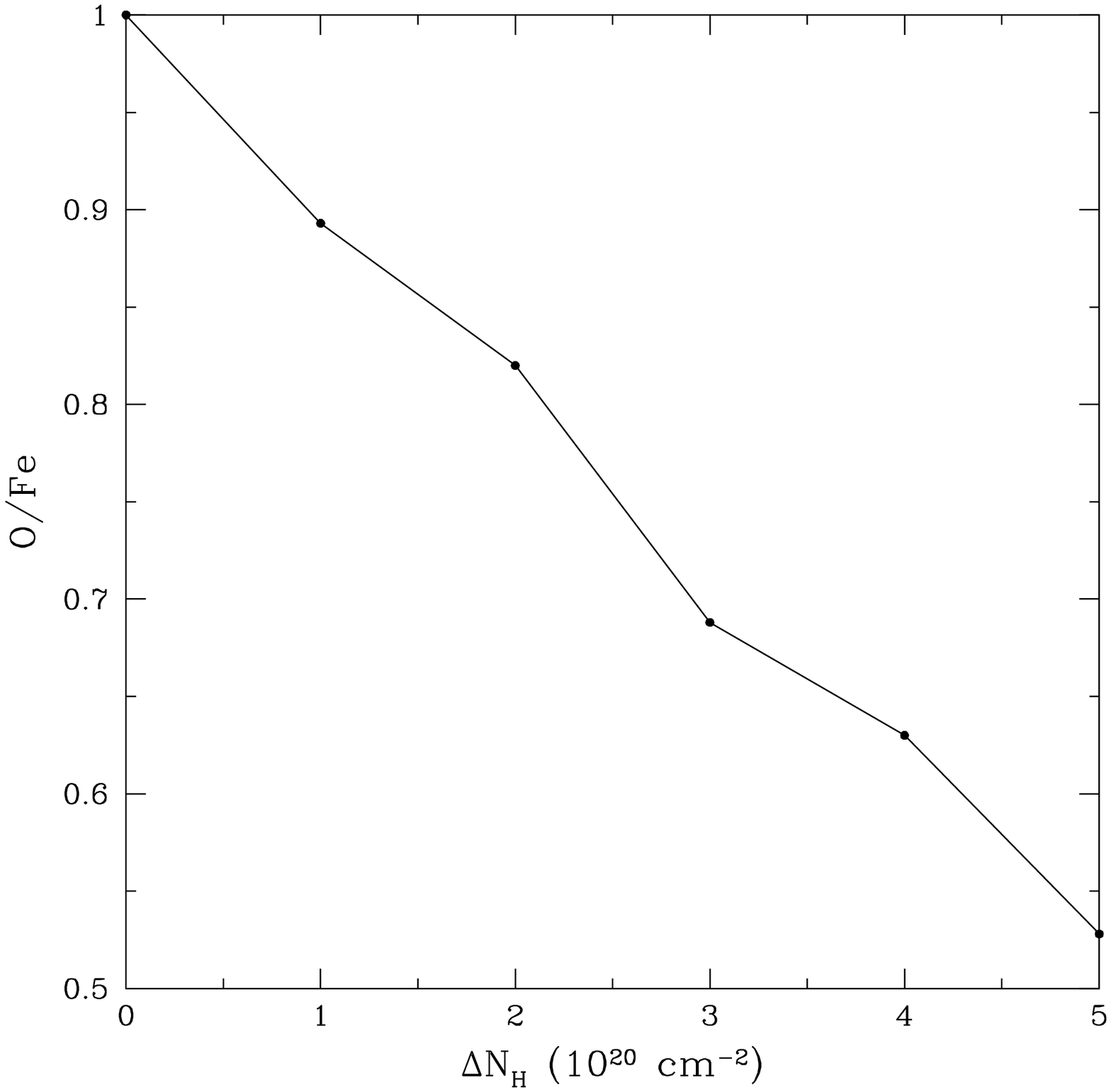}}
\caption{Best-fit O/Fe ratios derived from fitting an absorbed {\it apec} model 
with the absorption frozen at the galactic value to simulated
spectra generated from a {\it vapec} model with a solar O/Fe ratio and a range of 
excess absorption ($\Delta N_H$).}
\end{inlinefigure}

\begin{inlinefigure}
\center{\includegraphics*[width=1.00\linewidth,bb=19 145 583 706,clip]{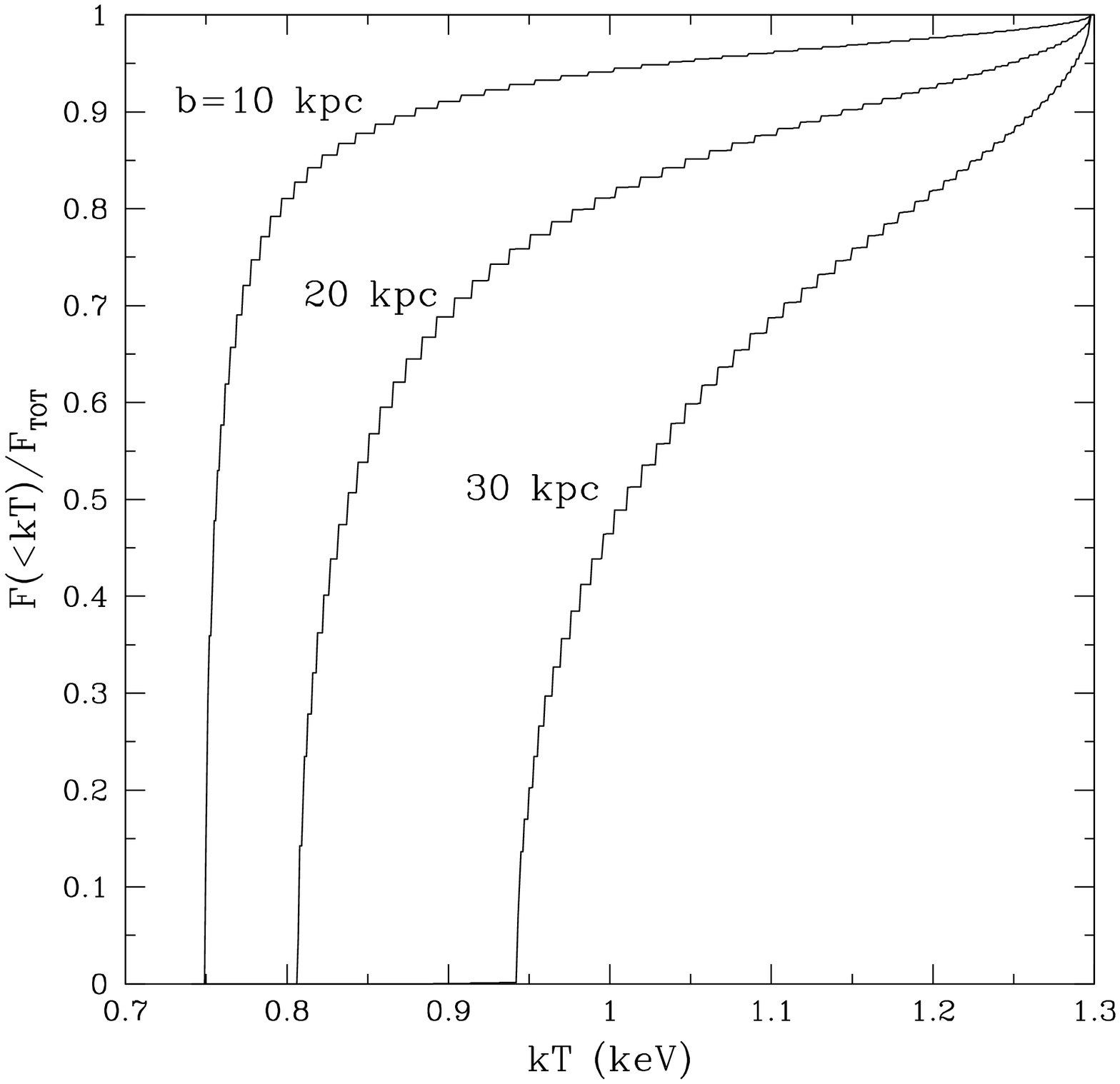}}
\caption{Cumulative fraction of the 0.5-3.0~keV X-ray flux, relative to the total flux, 
as a function of gas temperature along three different projected distances, b, from the 
center of NGC 5044.}
\end{inlinefigure}

\begin{inlinefigure}
\center{\includegraphics*[width=1.00\linewidth,bb=19 145 583 706,clip]{f10.eps}}
\caption{Scatter plot of the O/Fe ratios and Fe abundances for the 
high abundance clouds (black) and low abundance regions (red) derived from 
fitting the spectra to a two-temperature {\it vapec} model (see Table 5).  All error
bars are shown at $1 \sigma$.}
\end{inlinefigure}

\begin{inlinefigure}
\center{\includegraphics*[width=1.00\linewidth,bb=19 145 583 706,clip]{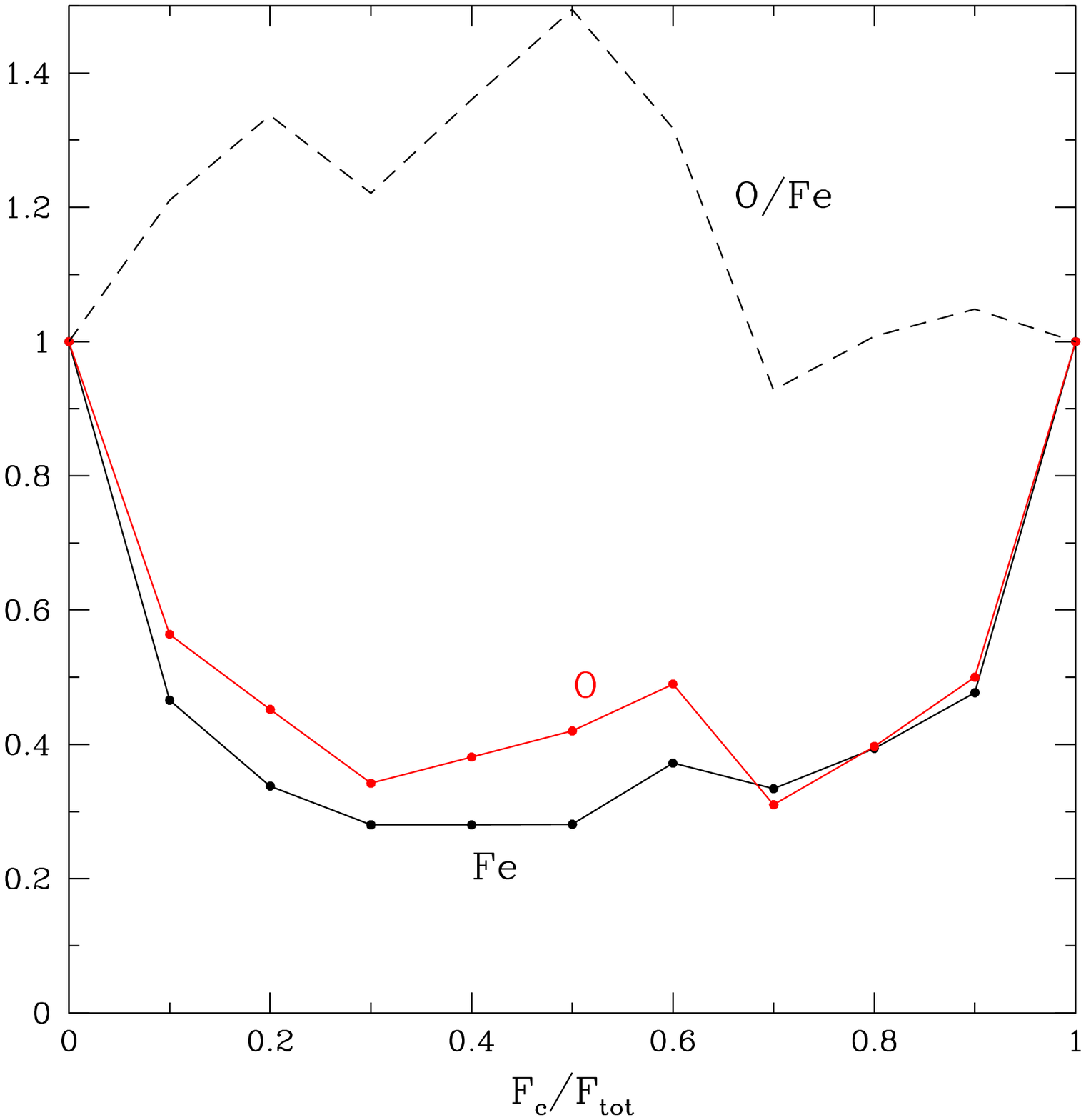}}
\caption{Best-fit values of O, Fe and O/Fe derived from fitting a single temperature model
to simulated two-temperature spectra with a lower temperature of 0.7~keV and an
upper temperature of 1.3~keV  and a range in flux ratios between the
two temperature components.  All simulated spectra have solar O and Fe abundances.}
\end{inlinefigure}

\begin{inlinefigure}
\center{\includegraphics*[width=1.00\linewidth,bb=19 145 583 706,clip]{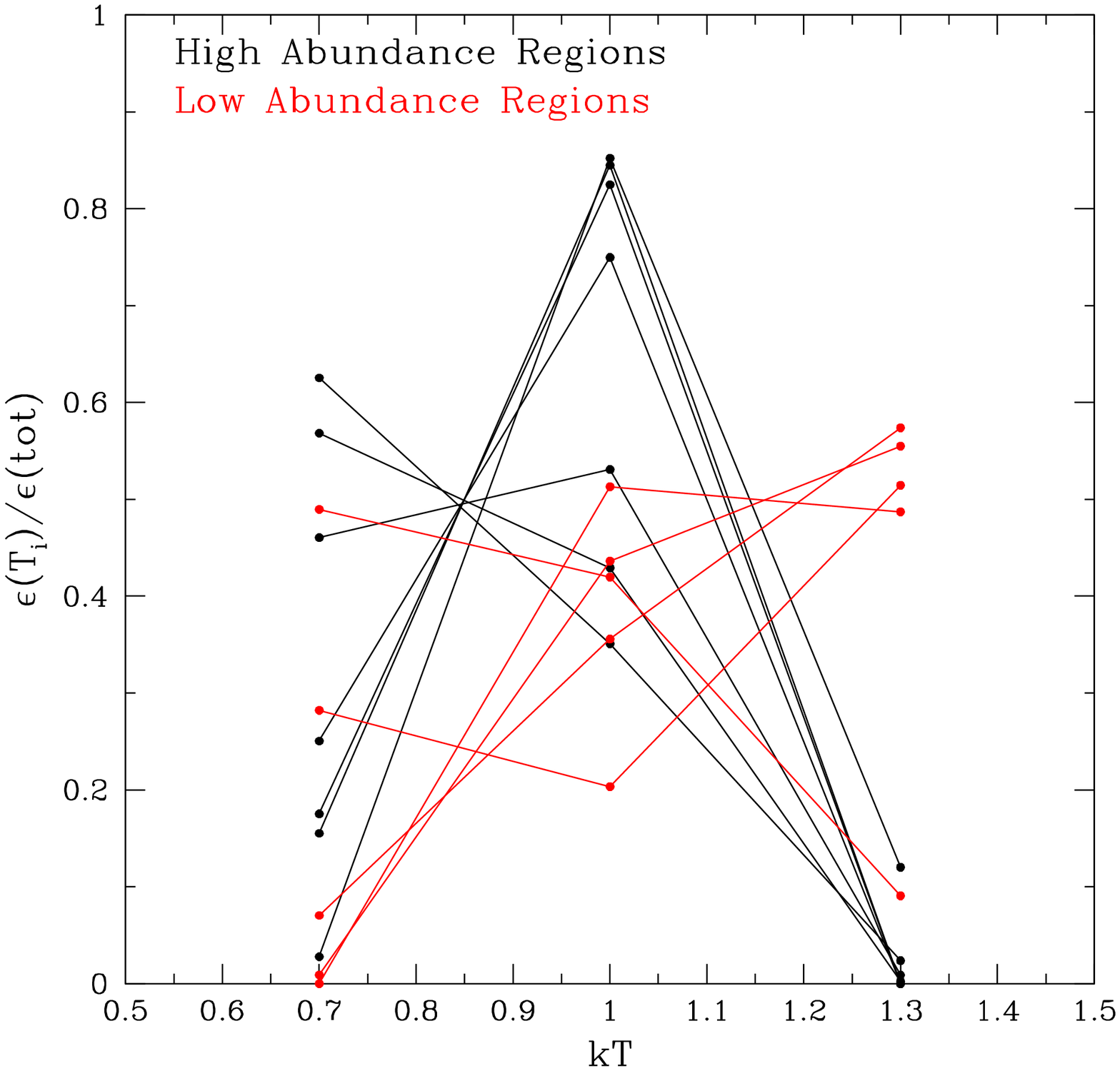}}
\caption{Fractional distribution of the emission measure derived from a three temperature fit
(kT=0.7, 1.0 and 1.3~keV) to the spectra from the high and low abundance regions.}
\end{inlinefigure}

\begin{inlinefigure}
\center{\includegraphics*[width=1.00\linewidth,bb=20 145 585 710,clip]{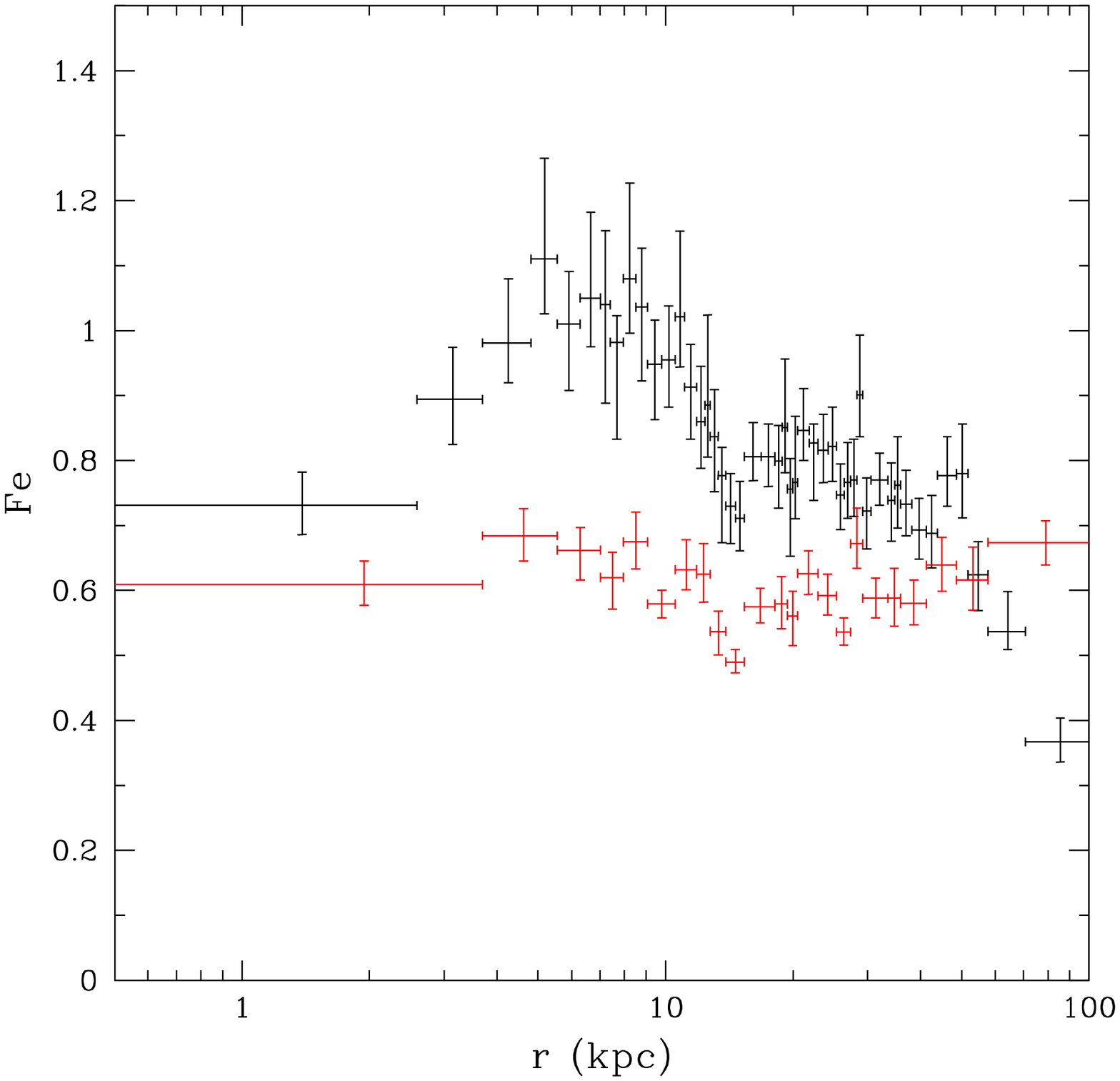}}
\caption{Projected Fe abundance profile relative to the solar value in
Grevesse \&  Sauval (1998).  All spectra were fit to absorbed
single temperature models {\it vapec} with $N_H$ frozen at the galactic value. 
The black data points are the best fit Fe abundances with all of the elemental 
abundances linked together.  The red data points are the best
fit Fe abundances with O allowed to vary independently of Fe. Spectra with 10,000 net
counts each were used to generate the black data points and spectra with 
20,000 net counts each were used to generate the red data points.}
\end{inlinefigure}

\begin{inlinefigure}
\center{\includegraphics*[width=1.00\linewidth,bb=20 145 589 695,clip]{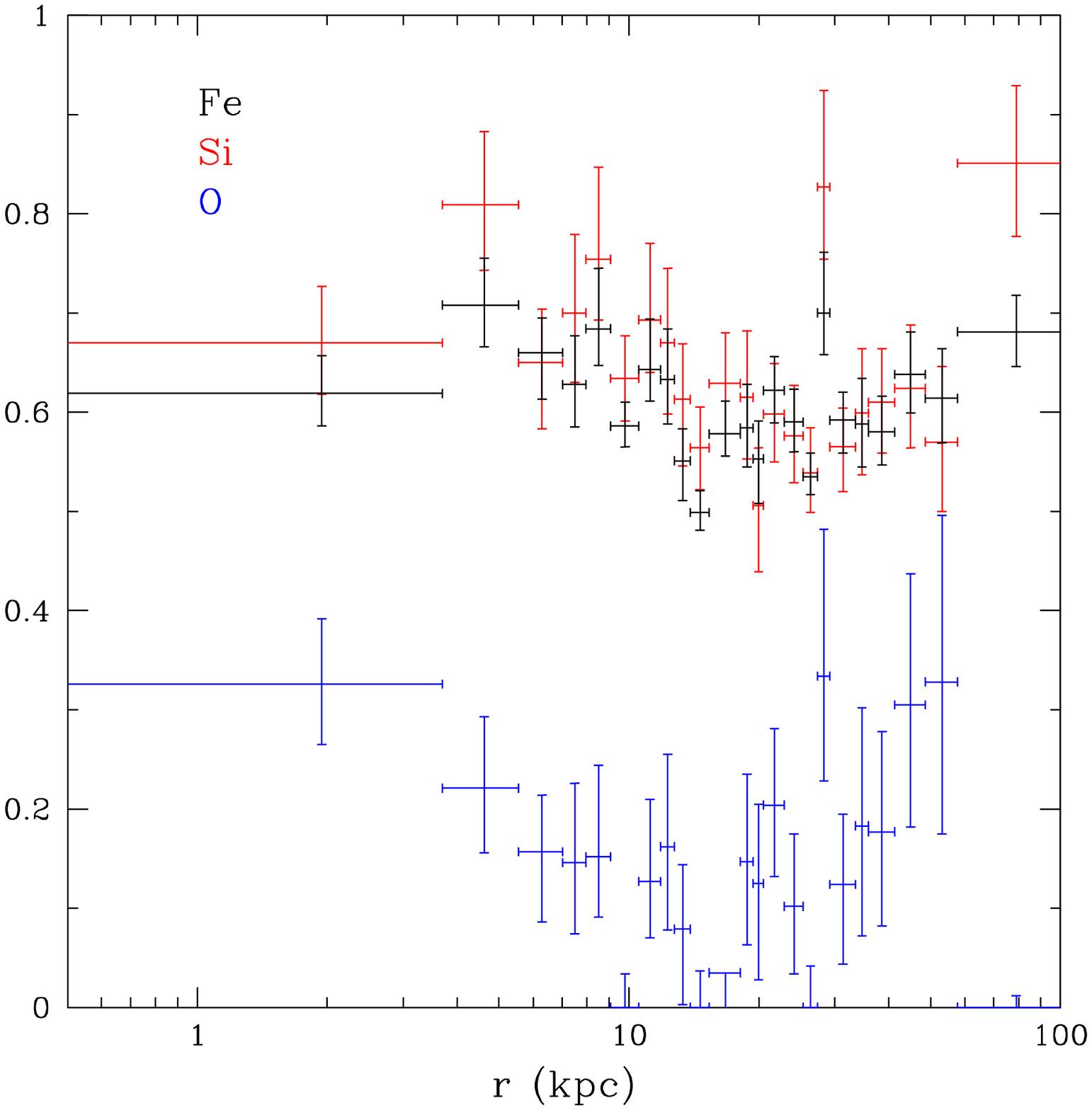}}
\caption{Projected O, Si and Fe abundance profiles relative to the solar values in
Grevesse \&  Sauval (1998).  All spectra have 20,000 net counts each and were fit to 
absorbed single temperature models ({\it vapec}) with $N_H$ frozen at the galactic value.}
\end{inlinefigure}

\begin{inlinefigure}
\center{\includegraphics*[width=1.00\linewidth,bb=20 145 585 695,clip]{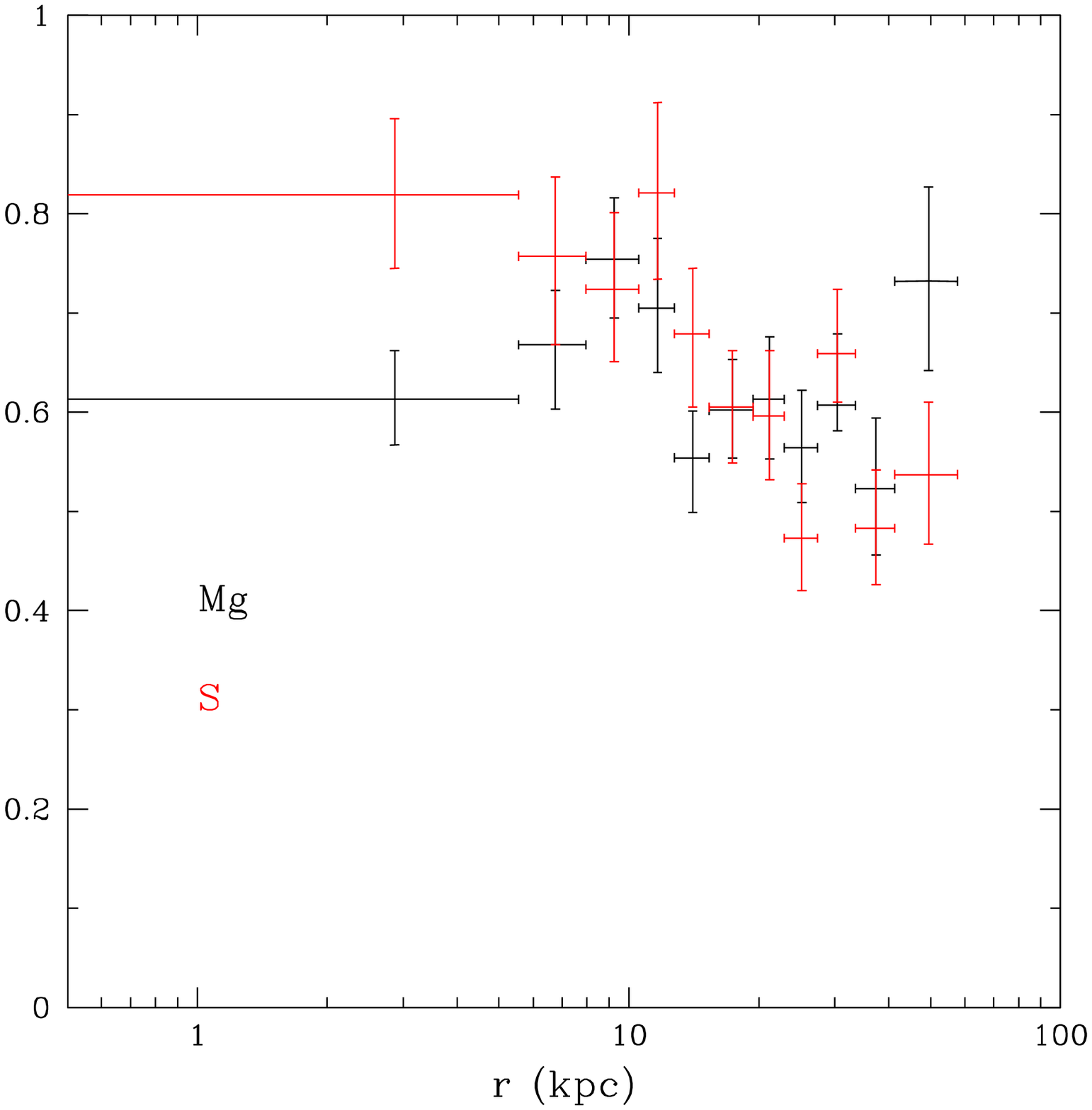}}
\caption{Projected Mg and S abundance profiles relative to the solar values in
Grevesse \&  Sauval (1998).  All spectra have 40,000 net counts each and were fit to 
absorbed single temperature models ({\it vapec}) with $N_H$ frozen at the galactic value.}
\end{inlinefigure}

\begin{inlinefigure}
\center{\includegraphics*[width=1.00\linewidth,bb=20 145 585 695,clip]{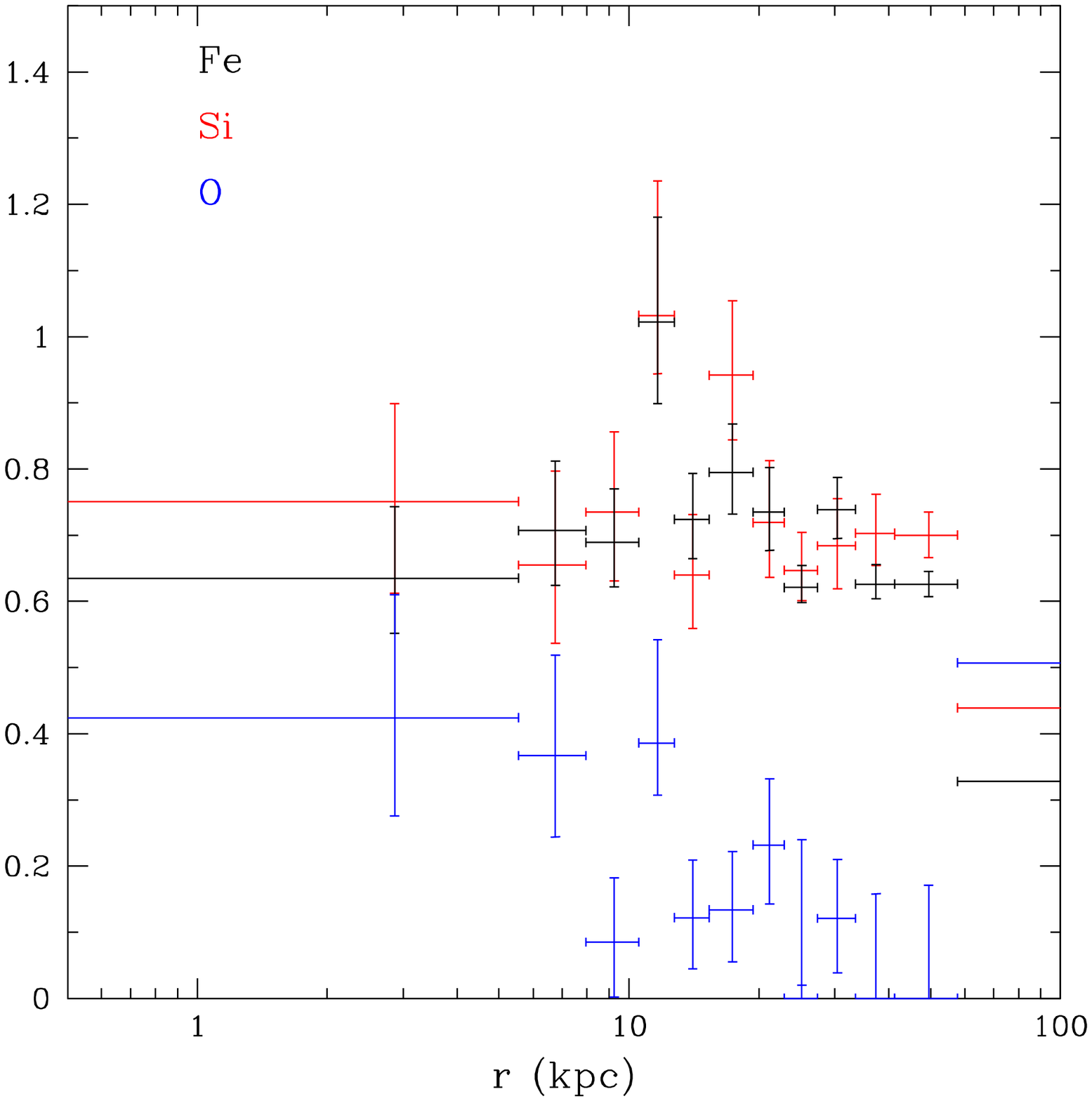}}
\caption{Deprojected Fe, Si and O abundance profiles relative to the solar values in
Grevesse \&  Sauval (1998).  All spectra have 40000 net counts each and were fit to 
absorbed single temperature models ({\it vapec}) with $N_H$ frozen at the galactic value.}
\end{inlinefigure}

\begin{inlinefigure}
\center{\includegraphics*[width=1.00\linewidth,bb=20 145 585 695,clip]{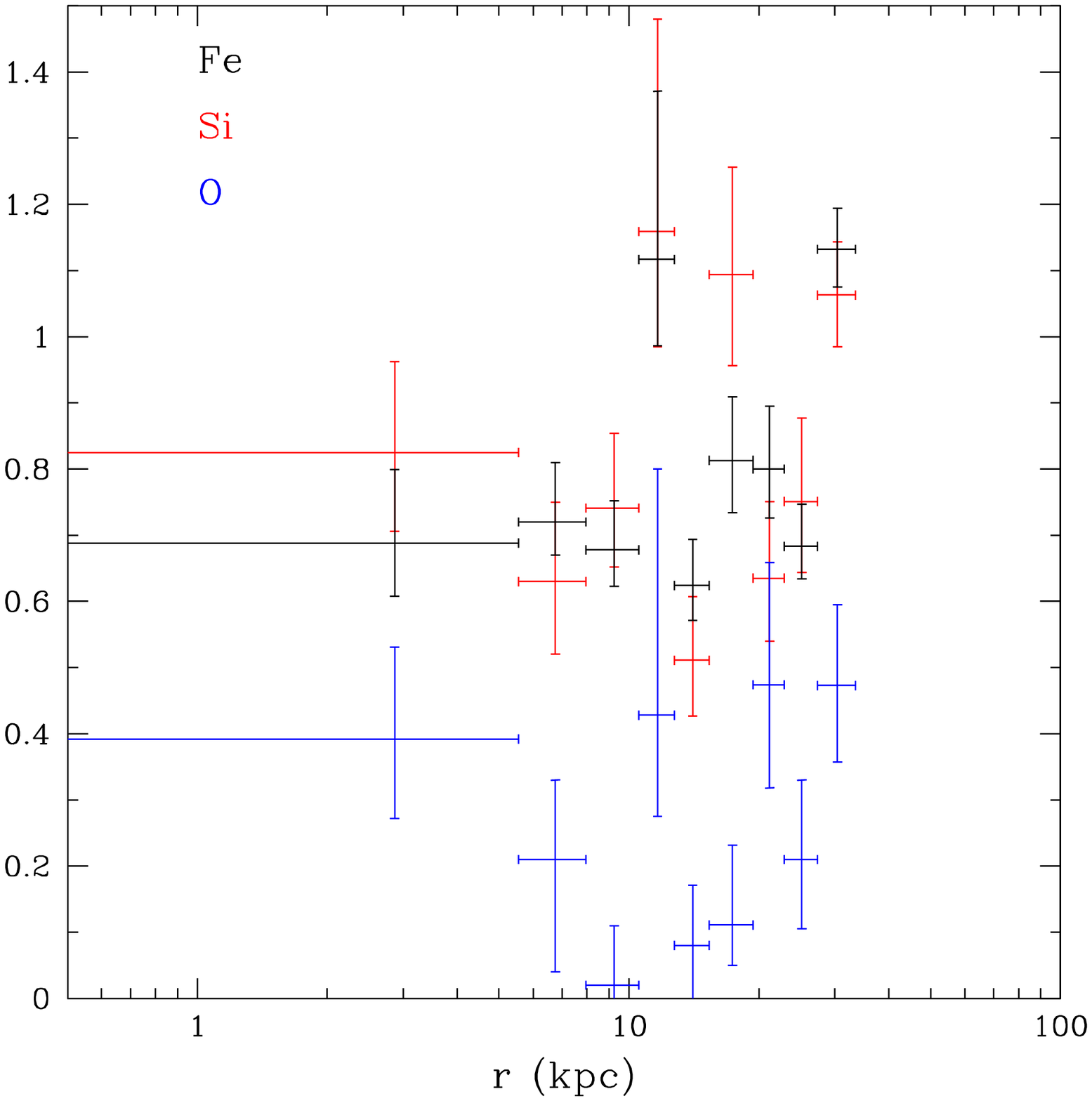}}
\caption{Deprojected Fe, Si and O abundance profiles relative to the solar values in
Grevesse \&  Sauval (1998).  All spectra have 40000 net counts each and were fit to 
absorbed two-temperature models with $N_H$ frozen at the galactic value.}
\end{inlinefigure}

\begin{inlinefigure}
\center{\includegraphics*[width=1.00\linewidth,bb=118 138 450 622,clip]{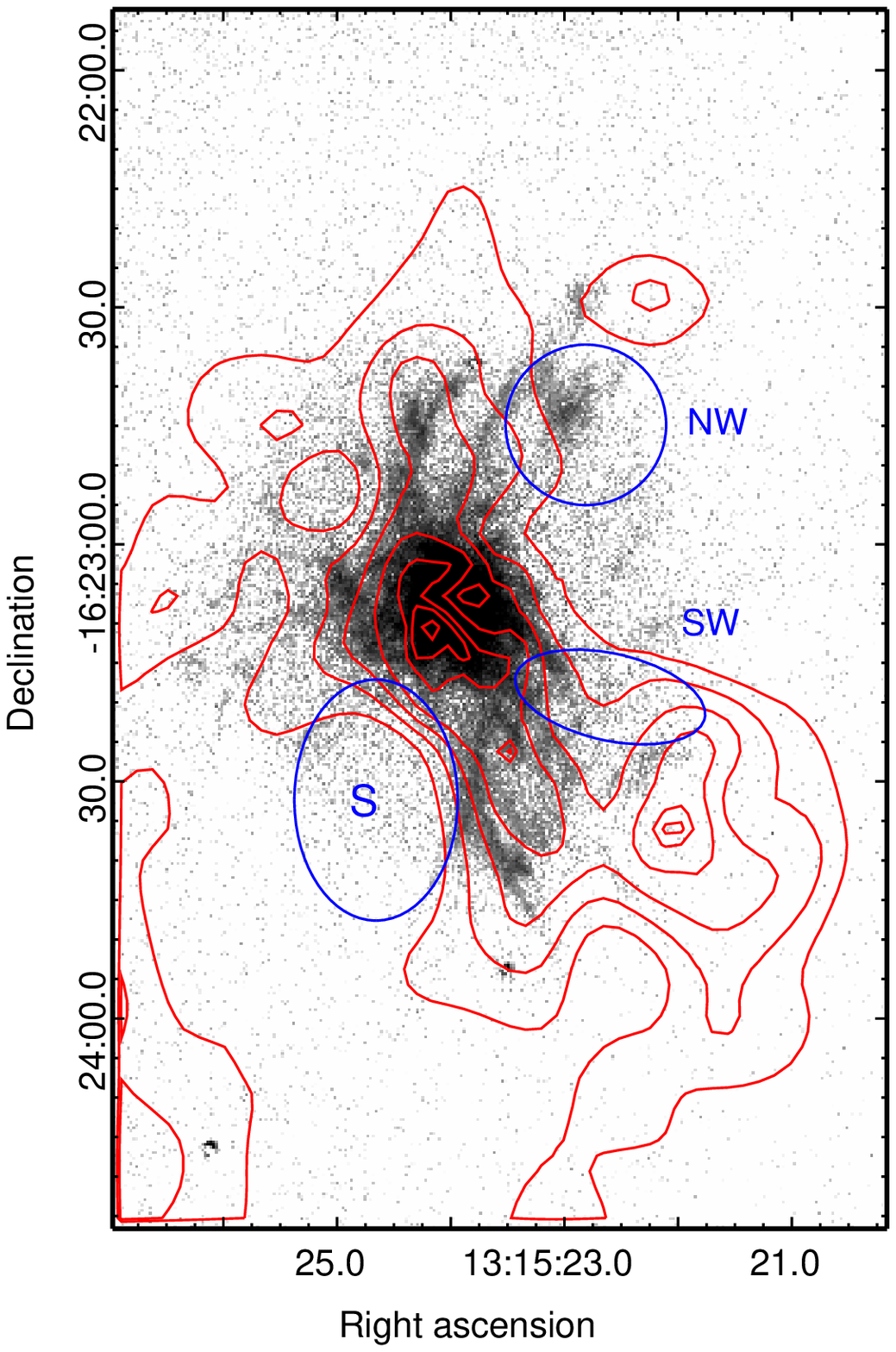}}
\caption{H$\alpha$ image along with temperature contours (red curves) derived from 
the temperature map shown in Figure 1, The gas temperatures increase from the smallest
to the largest contours. Also shown are the southern, northwestern and southwestern
cavities (blue curves).}
\end{inlinefigure}

\begin{inlinefigure}
\center{\includegraphics*[width=1.00\linewidth,bb=20 145 589 695,clip]{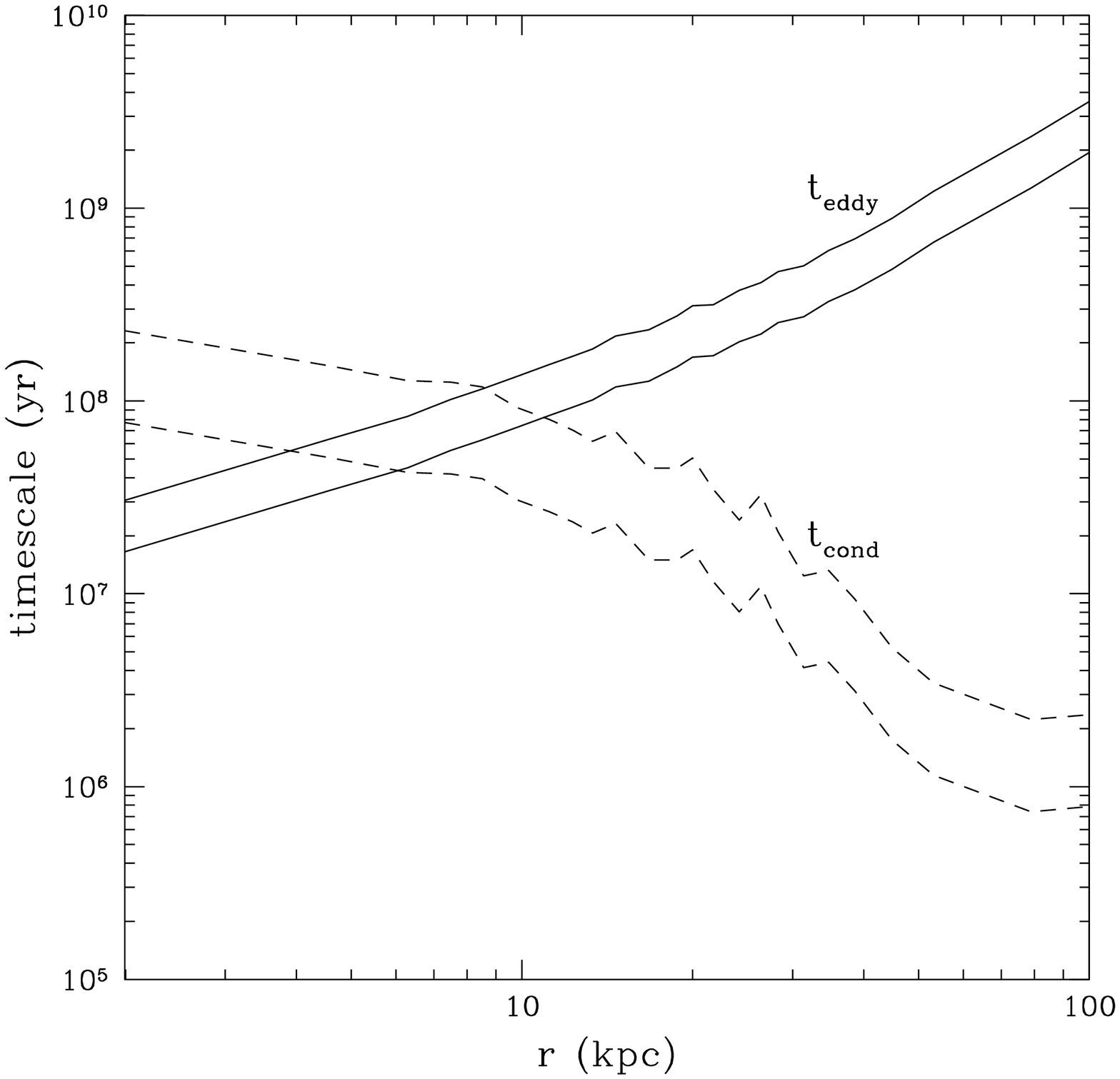}}
\caption{Eddy turnover time assuming that the dissipation of turbulent
kinetic energy locally balances radiative losses for $\alpha=0.2$ (lower solid curve) 
and $\alpha=0.5$ (upper solid curve). Conduction timescale ($t_{cond}$) to re-heat an 
embedded cool cloud to the ambient gas temperature assuming $f_s=0.3$ (lower dashed curve) 
and $f_s=0.1$ (upper dashed curve).}
\end{inlinefigure}

\begin{inlinefigure}
\center{\includegraphics*[width=1.00\linewidth,bb=20 145 589 695,clip]{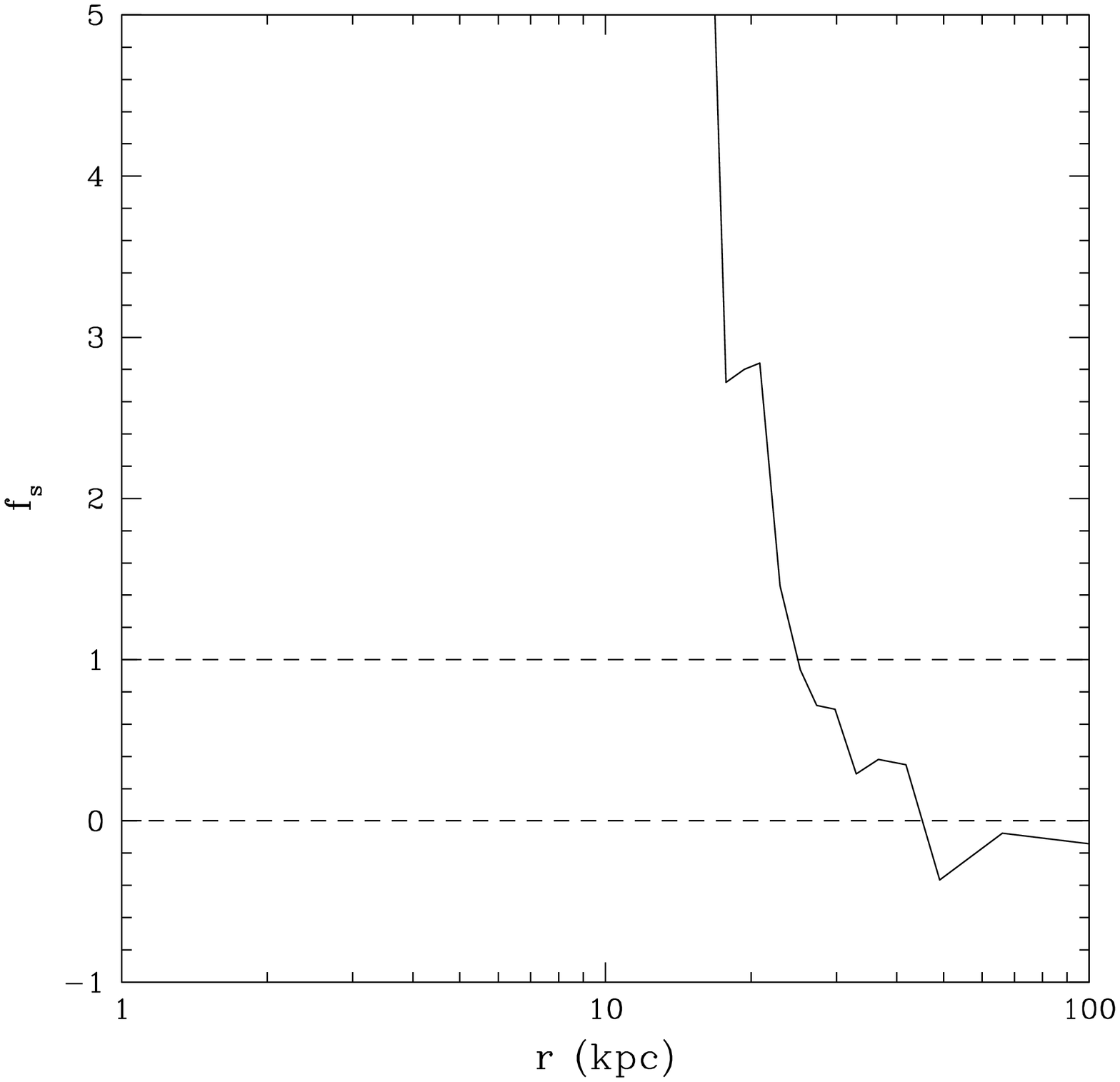}}
\caption{Ratio of the heat flux required to locally balance radiative losses to the
Spitzer value for the heat flux.}
\end{inlinefigure}

\newpage

\begin{table*}[t]
\begin{center}
\caption{Spectral Analysis Results using an absorbed {\it apec} Model}
\begin{tabular}{lccccc}
\hline
Region & kT & Z & $\chi^2$ & DOF & $\chi_{\nu}^2$  \\
& (keV) &  (solar) &  & & \\
\hline\hline
H1 & 1.01 (1.00-1.02) & 1.46 (1.23-1.65) &  109.7 &  90 & 1.22 \\
H2 & 0.93 (0.92-0.94) & 1.31 (1.15-1.46) &  201.2 & 103 & 1.95 \\
H3 & 0.82 (0.81-0.83) & 1.12 (1.05-1.25) &  282.0 & 118 & 2.39 \\
H4 & 0.89 (0.88-0.90) & 1.47 (1.19-1.88) &   83.4 &  71 & 1.17 \\
H5 & 0.79 (0.78-0.80) & 1.29 (1.13-1.50) &  157.5 & 100 & 1.57 \\
H6 & 0.78 (0.77-0.79) & 1.63 (1.39-2.57) &  105.0 &  74 & 1.42 \\
H7 & 0.93 (0.92-0.94) & 1.29 (1.02-1.57) &   71.1 &  64 & 1.11 \\
\hline
L1 & 0.81 (0.80-0.82) & 0.73 (0.69-0.81) &  154.3 & 109 & 1.42\\
L2 & 1.06 (1.05-1.07) & 0.55 (0.51-0.65) &   74.0 &  74 & 1.00 \\
L3 & 0.97 (0.95-0.99) & 0.39 (0.34-0.44) &  109.5 &  78 & 1.40 \\
L4 & 1.07 (1.05-1.09) & 0.53 (0.48-0.59) &  91.6  &  84 & 1.09 \\
L5 & 1.06 (1.05-1.07) & 0.54 (0.50-0.59) & 131.2  & 106 & 1.24 \\
\hline
\end{tabular}
\end{center}
\noindent
Notes: Spectral analysis results for the high and low abundance
regions highlighted in Figure 7. All spectra were fit in the 0.5-3.0~keV
energy band to a {\it phabs*apec} XSPEC spectral model with the absorption frozen
at the galactic value ($N_H=4.94 \times 10^{20}$~cm$^{-2}$) and
the Grevesse \&  Sauval (1998) abundance table. This table gives
the best-fit temperature (kT), abundance (Z), total $\chi^2$, 
degree of freedom (DOF) and $\chi_{\nu}^2$, along with the $1 \sigma$ errors.
\end{table*}

\begin{table*}[t]
\begin{center}
\caption{Spectral Analysis Results using an absorbed {\it vapec} Model}
\begin{tabular}{lcccccc}
\hline
Region & kT & Fe & O & $\chi^2$ & DOF  & F-Test Prob.\\
& (keV) & (solar) & (solar) & & &  \\
\hline\hline
H1 & 1.00 (0.99-1.01) & 0.92 (0.81-1.04) & 0.29 (0.11-0.50) &  96.4 &  89 &  $7.2 \times 10^{-4}$ \\
H2 & 0.91 (0.90-0.92) & 0.70 (0.66-0.74) & 0.00 ($< 0.05$)  & 129.8 & 102 &  $2.5 \times 10^{-11}$ \\
H3 & 0.80 (0.79-0.82) & 0.69 (0.65-0.74) & 0.08 (0.03-0.16) & 188.9 & 117 &  $8.3 \times 10^{-12}$ \\
H4 & 0.88 (0.86-0.89) & 0.73 (0.64-0.87) & 0.03 ($<0.23$)   &  76.4 &  70 &  $1.2 \times 10^{-4}$ \\
H5 & 0.78 (0.77-0.79) & 0.72 (0.67-0.81) & 0.10 (0.03-0.22) & 112.6 &  99 &  $8.9 \times 10^{-9}$ \\
H6 & 0.77 (0.76-0.78) & 0.95 (0.81-1.13) & 0.34 (0.17-0.57) &  91.0 &  73 &  $1.3 \times 10^{-3}$ \\  
H7 & 0.93 (0.91-0.94) & 1.08 (0.92-1.32) & 0.89 (0.52-1.43) &  70.8 &  63 &  0.61 \\
\hline
L1 & 0.80 (0.79-0.81) & 0.61 (0.57-0.66) & 0.26 (0.18-0.36) & 134.5 & 108 &  $1.2 \times 10^{-3}$ \\
L2 & 1.06 (1.05-1.07) & 0.53 (0.47-0.62) & 0.29 (0.17-0.56) &  73.7 &  73 &  0.59 \\
L3 & 0.97 (0.95-0.99) & 0.40 (0.34-0.46) & 0.30 (0.11-0.53) & 107.2 &  77 &  0.20 \\ 
L4 & 1.07 (1.05-1.09) & 0.50 (0.44-0.56) & 0.30 (0.09-0.52) &  89.5 &  83 &  0.17 \\
L5 & 1.06 (1.05-1.09) & 0.51 (0.46-0.56) & 0.27 (0.11-0.42) & 128.3 & 105 &  0.13 \\
\hline
\end{tabular}
\end{center}
\noindent
Notes:  All spectra were fit in the 0.5-3.0~keV energy band to a {\it phabs*vapec} XSPEC 
spectral model with the absorption frozen at the galactic value and the O and
Fe abundances treated as free parameters. See the notes to Table 1 for further details.
\end{table*}

\begin{table*}[t]
\begin{center}
\caption{Spectral Analysis Results Using an Absorbed {\it apec} Model in the 0.8-3.0~keV Band}
\begin{tabular}{lcccc}
\hline
Region & kT & Z & $\chi^2$ & DOF  \\
& (keV) &  (solar) &  \\
\hline\hline
H1 & 1.01 (1.00-1.02) & 1.04 (0.91-1.32) &  84.8  & 73  \\
H2 & 0.91 (0.90-0.93) & 0.70 (0.62-0.80) &  94.3  & 83  \\
H3 & 0.80 (0.79-0.81) & 0.63 (0.58-0.69) & 154.9  & 98  \\
H4 & 0.88 (0.87-0.89) & 0.73 (0.58-0.93) &  48.5  & 59  \\
H5 & 0.78 (0.77-0.79) & 0.73 (0.65-0.85) &  91.1  & 80  \\
H6 & 0.76 (0.75-0.77) & 0.97 (0.82-1.41) &  70.6  & 59  \\
H7 & 0.93 (0.92-0.94) & 1.17 (0.89-1.62) &  58.6  & 52  \\
\hline
L1 & 0.81 (0.80-0.82) & 0.59 (0.53-0.64) & 111.7 &  89 \\
L2 & 1.06 (1.05-1.07) & 0.47 (0.40-0.54) &  68.2  & 61  \\ 
L3 & 0.99 (0.97-1.01) & 0.39 (0.29-0.43) &  70.1  & 62  \\ 
L4 & 1.05 (1.04-1.06) & 0.36 (0.31-0.42) &  69.7  & 69 \\ 
L5 & 1.07 (1.06-1.08) & 0.52 (0.47-0.59) & 102.3  & 86 \\ 
\hline
\end{tabular}
\end{center}
\noindent
Notes:  All spectra were fit in the 0.8-3.0~keV energy band to a {\it phabs*apec} XSPEC 
spectral model with the absorption frozen at the galactic value.
See the notes to Table 1 for further details.
\end{table*}

\begin{table*}[t]
\begin{center}
\caption{Spectral Analysis Results Using using an Absorbed {\it vapec} Model with Free Absorption}
\begin{tabular}{lccccccc}
\hline
Region & kT & Fe & O & $\rm{N_H}$ & $\chi^2$ & DOF  & F-Test Prob.\\
& (keV) & (solar) & (solar) & $(10^{20} cm^2)$ &  &  & \\
\hline\hline
H1 & 1.00 (0.99-1.01) & 0.92 (0.81-1.04) & 0.27 (0.06-0.54) & 4.73 (2.98-6.54) (1.92-7.74) &  96.4 &  88  & 1.00 \\
H2 & 0.91 (0.90-0.92) & 0.67 (0.63-0.73) & 0.00 ($<0.05$)  & 6.09 (5.39-7.18) (4.10-7.94)  & 129.0 & 101  & 0.43 \\
H3 & 0.79 (0.78-0.80) & 0.71 (0.66-0.76) & 0.19 (0.11-0.28) & 7.62 (6.73-8.53) (6.17-9.12) & 179.4 & 116  & 0.010 \\
H4 & 0.84 (0.83-0.85) & 0.73 (0.62-0.88) & 0.12 ($<0.42$)  & 9.36 (7.19-11.7) (5.74-13.4)  &  63.3 &  69  & $3.3 \times 10^{-4}$ \\
H5 & 0.77 (0.76-0.78) & 0.76 (0.68-0.86) & 0.22 (0.09-0.37) & 7.24 (5.87-8.62) (5.00-9.63) & 109.8 &  98  & 0.11 \\
H6 & 0.76 (0.75-0.77) & 0.98 (0.84-1.40) & 0.45 (0.24-1.07) & 7.13 (5.90-9.95) (3.75-11.7) &  90.0 &  72  & 0.37 \\
H7 & 0.93 (0.91-0.95) & 1.07 (0.88-1.33) & 0.82 (0.41-1.47) & 4.22 (1.71-6.92) (1.94-8.70) &  70.7 &  62  & 0.88 \\
\hline
L1 & 0.80 (0.79-0.81) & 0.64 (0.59-0.69) & 0.37 (0.25-0.50) & 6.90 (5.80-7.97) (5.12-8.02) & 131.2 & 107 & 0.10 \\
L2 & 1.05 (1.04-1.07) & 0.55 (0.48-0.65) & 0.46 (0.20-0.88) & 7.14 (5.04-9.60) (3.55-11.4) &  72.7 &  72  & 0.33 \\
L3 & 0.95 (0.93-0.97) & 0.41 (0.34-0.48) & 0.54 (0.24-0.86) & 8.25 (5.88-10.6) (3.45-4.86) & 105.3 &  76 & 0.24 \\
L4 & 1.05 (1.03-1.07) & 0.52 (0.46-0.64) & 0.72 (0.41-1.25) & 10.7 (8.53-13.2) (6.90-15.4) &  83.3 &  82 & 0.015 \\
L5 & 1.06 (1.04-1.08) & 0.51 (0.46-0.56) & 0.35 (0.16-0.55) & 6.13 (4.60-7.68) (3.64-8.76) & 127.8 & 104 &  0.51 \\
\hline
\end{tabular}
\end{center}
\noindent
Notes:  All spectra were fit in the 0.5-3.0~keV energy band to a {\it phabs*vapec} XSPEC 
spectral model with free absorption.  See the notes to Table 1 for further details.
Both $1 \sigma$ and 90\% uncertainties are given for $N_{\rm H}$.
\end{table*}

\newpage

\begin{table*}[t]
\begin{center}
\caption{Spectral Analysis using a two-temperature {\it vapec} model}
\begin{tabular}{lcccccccc}
\hline
Region & $\rm{kT_c}$ & $\rm{kT_h}$ & Fe & O & $\rm{F_c/F_{tot}}$ & $\chi^2$ & DOF  & F-Test Prob.\\
& (keV) & (keV) & (solar) & (solar) & & & & \\
\hline\hline
H1 & 1.00 (0.99-1.01) &  -               & 0.92 (0.81-1.04) & 0.27 (0.06-0.54) & -    &   -    &  - & - \\
H2 & 0.83 (0.82-0.85) & 1.22 (1.11-1.33) & 0.87 (0.78-0.96) & 0.00 ($< 0.07)$  & 0.75 &  115.0 & 100 & $2.40 \times 10^{-3}$ \\
H3 & 0.78 (0.77-0.79) & 1.36 (1.24-1.47) & 1.00 (0.90-1.06) & 0.29 (0.21-0.34) & 0.89 &  154.9 & 115 & $1.12 \times 10^{-5}$ \\
H4 & 0.80 (0.52-0.85) & 1.00 (0.90-1.08) & 0.95 (0.78-1.17) & 0.06 ($< 0.27)$) & 0.88 &   57.7 &  68 & $7.37 \times 10^{-5}$ \\
H5 & 0.77 (0.76-0.77) & 2.30 (1.76-3.23) & 1.08 (0.98-1.25) & 0.36 (0.25-0.66) & 0.93 &  100.2 &  97 & $3.57 \times 10^{-3}$ \\
H6 & 0.74 (0.67-0.76) & 1.30 (0.96-1.57) & 1.51 (1.04-2.40) & 0.78 (0.49-1.29) & 0.90 &  85.1  &  71 & 0.092 \\   
H7 & 0.93 (0.91-0.95) & -                & 1.07 (0.88-1.33) & 0.82 (0.41-1.47) & -    &  -     &  -  & - \\
\hline
L1 & 0.76 (0.74-0.77) & 1.21 (1.10-1.35) & 0.98 (0.86-1.13) & 0.60 (0.45-0.77) & 0.78 &  91.5  & 106 & $1.35 \times 10^{-9}$ \\  
L2 & 0.97 (0.91-1.02) & 1.36 (1.20-1.71) & 0.85 (0.70-1.24) & 0.71 (0.42-1.18) & 0.60 &  67.5  &  71 & 0.044 \\
L3 & 0.71 (0.63-0.79) & 1.23 (1.15-1.37) & 0.78 (0.71-1.00) & 0.62 (0.23-0.98) & 0.40 &  82.0  &  75 & $1.61 \times 10^{-4}$ \\
L4 & 1.07 (1.05-1.09) & -                & 0.50 (0.44-0.56) & 0.30 (0.09-0.52) & -    &   -    &  -  &  \\
L5 & 0.62 (0.27-0.71) & 1.16 (1.12-1.19) & 0.77 (0.65-0.86) & 0.46 (0.16-0.64) & 0.13 & 110.2  &  103 & $3.97 \times 10^{-4}$ \\
\hline
\end{tabular}
\end{center}
Notes:  All spectra were fit in the 0.5-3.0~keV energy band to a {\it phabs*(vapec+vapec)} XSPEC 
spectral model with the absorption frozen at the galactic value.  
The lower best-fit temperature is given by $\rm{kT_c}$ and the upper best-fit temperature
is given by $\rm{kT_h}$.  The ratio between the fluxes in the cooler spectral component to
the total flux in the 0.5-3.0~keV energy band is given by $\rm{F_c/F_{tot}}$.
See the notes to Table 1 for further details.
\end{table*}

\end{document}